\documentclass[11pt]{article}

\usepackage[utf8]{inputenc}
\usepackage{graphicx}
\usepackage{amsmath,amsfonts, amssymb}

\usepackage{cite}

\usepackage{multirow}
\usepackage{graphicx}
\usepackage{hhline}
\usepackage[shortlabels]{enumitem}
\usepackage{hyperref}
\hypersetup{breaklinks=true, colorlinks=false, linkcolor=black, urlcolor=black}
\usepackage{breakurl}
\usepackage{setspace}
\allowdisplaybreaks

\usepackage{xcolor}

\begin{document}

\title{
\LARGE
Interplay between Electroweak Symmetry
Breaking and Higgs Portal Dark Matter}

\author{
  Sreemanti Chakraborti,$^{(1)}$\thanks{E-mail:
    \tt sreemanti.chakraborti@durham.ac.uk}
  \
  Andr\'e Milagre,$^{(2)}$\thanks{E-mail:
    \tt andre.milagre@tecnico.ulisboa.pt}
  \
  Rui Santos,$^{(3,4)}$\thanks{E-mail:
    \tt rasantos@fc.ul.pt}
  \
  and Jo\~ao P. Silva$^{(2)}$\thanks{E-mail:
    \tt jpsilva@cftp.ist.utl.pt}
  \\*[3mm]
  $^{(1)}\!$
  \small Institute for Particle Physics Phenomenology, Department of Physics, \\
  \small Durham University, Durham, DH1 3LE, United Kingdom
  \\*[2mm]
  $^{(2)}\!$
  \small Instituto Superior T\'ecnico, Universidade de Lisboa, \\
  \small Avenida Rovisco Pais 1, 1049-001 Lisboa, Portugal
  \\*[2mm]
  $^{(3)}\!$
  \small Centro de F\'isica T\'eorica e Computacional, Faculdade de Ci\^encias, \\
  \small  Universidade de Lisboa, Campo Grande, Edif\'icio C8 1749-016 Lisboa, Portugal
  \\*[2mm]
  $^{(4)}\!$
  \small ISEL-Instituto Polit\'ecnico de Lisboa 1959-007 Lisboa, Portugal
}

\date{\today}

\maketitle

\begin{abstract}
Models of Dark Matter must contend with the fact that
the presence of electroweak symmetry breaking along the thermal evolution of the Universe
modifies the masses, interactions and, thus, the thermally averaged cross sections.
We study in detail the impact of taking (not taking) the presence of the electroweak symmetry
breaking into account in the calculations of the Dark Matter relic density
in Higgs portal models,
providing a model-independent measure of such differences.
By focusing on a particular model,
we show that ignoring this effect can lead to the inclusion (exclusion) of
wrong (viable) regions of parameter space.
\end{abstract}

%
\section{Introduction}
The spectrum of temperature fluctuations in the Cosmic Microwave Background has offered physicists an extraordinary opportunity to measure the relative abundances of the various energy components of the Universe.
The structures that make up these fluctuations are consistent with the existence of a non-baryonic matter component that exceeds the amount of ordinary baryonic matter by a factor around five, offering, together with a number of experimental results from many other experimental sources \cite{Cirelli:2024ssz}, compelling evidence for the existence of Dark Matter (DM).
The best determination of the present DM relic density is~\cite{Planck:2018vyg}
\begin{equation}
	\Omega h^2|_{\rm Obs} = 0.1200 \pm 0.0012,
	\label{obs_relic}
\end{equation}
where $\Omega$ is the current energy density in the form of
DM relative to the critical energy density of the Universe,
and $h$ is today’s reduced Hubble constant in the units of $100\,$km/s/MPc.

Amongst the plethora of DM production mechanisms found in the literature~\cite{DEramo:2010keq, Hochberg:2014dra, Kuflik:2015isi, DAgnolo:2015ujb, Pappadopulo:2016pkp, Garny:2017rxs, DAgnolo:2020mpt, Smirnov:2020zwf, Fitzpatrick:2020vba, Kramer:2020sbb, Hryczuk:2021qtz, Bringmann:2021tjr}, the thermal freeze-out scenario stands out for mirroring the thermal behavior of particles in the Standard Model (SM)~\cite{Bauer:2017qwy}.\footnote{Namely, photons and neutrinos.} This mechanism assumes that in the early stages of our Universe, \textit{i.e.} at very high temperatures, the interaction rate between DM and particles in the primordial thermal bath was strong enough to keep them in chemical and thermal equilibrium. \textcolor{black}{However, as the Universe expanded and cooled, the annihilation rate of DM into lighter particles gradually became subdominant to the Hubble expansion rate, resulting in the decoupling of DM from the thermal plasma and the subsequent \textit{freeze-out} of its comoving number density.}

In the freeze-out scenario, the present relic density is computed by solving the Boltzmann Equation, which describes the change in the number density of the DM species over time. As an input, one needs to provide a cosmological history for our Universe (contained in the Hubble parameter) and a set of thermally averaged DM depletion processes (coming from the particle physics model chosen, provided it has a DM candidate). \textcolor{black}{The solution to the Boltzmann Equation can then be rewritten as an integral over temperature, starting from an initial epoch where dark matter is in chemical equilibrium with a radiation-dominated thermal bath and extending until its comoving abundance approaches a constant asymptotic value.}
With this information, one may predict the corresponding DM relic density and test it against the observed value in Eq.~\eqref{obs_relic}.

In practice, analytical and numerical solutions are hard to find, so one usually resorts to approximations to perform the aforementioned integral. One such approximation relies on the assumption that DM particles remain in thermal equilibrium with the primordial bath until the freeze-out instant, implying that the Boltzmann equation becomes dominated by DM depletion processes occurring at temperatures below freeze-out, and hence below the DM mass scale. In effect, this reduces the problem to integrating the Boltzmann Equation only from freeze-out to the present time~\cite{Cline:2013gha,Arcadi:2024ukq}. Since the relevant integration range lies at lower temperatures, the process becomes effectively low-energy, and finite temperature corrections are typically neglected in this scenario. This contrasts with other DM production mechanisms where these higher-order corrections play an important role in accurately predicting the present DM abundance\textcolor{black}{\cite{Baker:2017zwx,Chakrabarty:2022bcn,Biondini:2020ric,Biondini:2023hek,Becker:2025lkc}}. 

{\color{black}
The \textit{standard} approach, therefore, is to calculate all thermally averaged DM depletion processes in the low-energy regime and input them into the Boltzmann equation. However, this reasoning overlooks the fact that the Higgs boson acquires a non-zero vacuum expectation value (vev), triggering electroweak symmetry breaking (EWSB) at temperatures around $T_{\rm EWSB} \approx 160$\,GeV~\cite{Laine:2015kra}.
Consequently, particle masses and interaction rates must change across the EWSB transition.
In this sense, we propose an \textit{improved} approach for calculating the DM relic density, in which interactions are evaluated both before and after EWSB.}

As we will demonstrate, neglecting the effects of EWSB in the calculation of the DM relic density is especially concerning \textcolor{black}{for Higgs portal models where} DM freezes out before EWSB, \textit{i.e.} for DM candidates with a mass above $\sim 4$\,TeV, and can lead to large discrepancies between the \textit{standard} and \textit{improved} approaches.
We prove that the deviations from the \textit{standard} approach are driven by a change in the freeze-out instant and/or by a change in the thermally averaged DM depletion processes. In addition, we provide an analytical way to quantify this deviation.

{\color{black}
Several well-known departures from the standard freeze-out picture have been explored in the literature.
For instance, the possibility that particle masses may vary over the history of the Universe has been explored in:
Variable-Mass particle models, where a scalar DM candidate is coupled to a quintessence field ~\cite{Anderson:1997un,Rosenfeld:2005pw,Franca:2003zg}; 
spontaneous freeze-out, where fermionic DM acquires its mass through an early phase transition, and freezes out due to its mass becoming larger than the plasma temperature~\cite{Heurtier:2019beu};
the Flip-Flop vev mechanism, where an auxiliary scalar field develops a temporary vacuum expectation value, inducing a period where DM is allowed to decay~\cite{Baker:2018vos};
the context of two-step phase transitions, where thermal masses influence the dynamical freeze-in production of DM, primarily due to the kinematical thresholds in decay and scattering processes~\cite{Bian:2018bxr};
supercool DM, where the Universe undergoes a brief period of thermal inflation during which all particles become massless~\cite{Hambye:2018qjv} or where it undergoes a confining phase transition, resulting in the formation of massive DM bound states~\cite{Baldes:2021aph}.
Other detours from the standard freeze-out include resonant or co-annihilation behaviour~\cite{Griest:1990kh,Edsjo:1997bg}, 
departures from kinetic equilibrium~\cite{Bringmann:2006mu,Binder:2017rgn}, 
and finite-temperature corrections to particle masses and couplings~\cite{Baker:2017zwx,Chakrabarty:2022bcn}.
Each represents an independent refinement of the basic framework.
The effect considered in this work—associated with the two-phase electroweak structure—is of a different and complementary nature: it arises from a change in the field content across the electroweak transition rather than from higher-order thermal corrections within a single phase.
The distinction between the \textit{standard} and \textit{improved} approaches is therefore \textbf{structural}: whether freeze-out occurs in the unbroken or broken electroweak phase determines which degrees of freedom and interaction channels are available, and hence which annihilation processes contribute to the relic density.}

Most importantly, the inclusion of DM depletion processes before and after EWSB has also been considered in a model with a light scalar DM candidate produced via freeze-in~\cite{Heeba:2018wtf}, and in a two-component DM model where one particle freezes in and the other freezes out~\cite{Bhattacharya:2021rwh}.
However, to the best of our knowledge, a systematic comparison between the \textit{standard} and \textit{improved} approaches and its physical implications is lacking in the literature.

The paper is organized as follows.
In Section~\ref{sec:relic}, we start by discussing the 
\textit{standard} way of computing thermal
relic densities performed using the broken phase model throughout the thermal evolution of the Universe.
Then, we repeat the calculation using the unbroken phase model
before EWSB, and the broken phase model after EWSB, 
which we dub the \textit{improved} approach.
Figures of merit aimed at comparing the two approaches are introduced in
Subsection~\ref{subsec:comparison}.
In Section~\ref{sec:casestudy}, we introduce a simple model illustrating
the differences between the two approaches, which are quantitatively explored in
Section~\ref{sec:results}.
The conclusions appear in Section~\ref{sec:conclusion}.
Several detailed aspects of the calculation have been relegated to the appendices.

%
\section{Relic density: standard vs. improved approaches}
\label{sec:relic}

\subsection{Relic density: the standard approach}
\label{subsec:standard}

Consider a SM extension with a single DM candidate of mass $m_{\rm DM}$ which, at high enough temperatures, is kept in thermal and chemical equilibrium with the primordial bath through $2\to2$ scattering processes.
To calculate the present abundance of DM in our Universe, we make use of the Boltzmann Equation to track its number density $n_{\rm DM}$, over time.
For practical purposes, it is useful to define the particle number density per comoving volume (or yield) as $Y_{\rm DM}\equiv n_{\rm DM}/s$, where $s$ is the entropy density of the Universe.
The Boltzmann equation describing the evolution of the yield in this scenario can be written as~\cite{Cline:2013gha,Steigman:2012nb}
\begin{equation}
	\frac{dY_{\rm DM}}{dy}=Z(y)
	\left(\overline{Y}(y)^2-Y_{\rm DM}^2\right),
    \label{eq:beq_standard}
\end{equation}
where\footnote{
Usually, the independent variable 
is defined as the ratio of the 
DM mass to the temperature of the Universe, \textit{i.e.}
$x= m_{\rm DM} / T$.
However, since we consider a scenario where $m_{\rm DM}$ 
changes discontinuously across a phase transition, 
this definition is ill-defined, and we work with $y =1/T$ instead.}
$y\equiv 1/T$
is the ratio of the DM mass to the temperature of the Universe, 
\begin{equation}
    \overline{Y}(y) \equiv \frac{45}{4 \pi^4} \frac{m_{\rm DM}^2 y^2}{h_{\mathrm{eff}}(y)} 
    K_2 \left(m_{\rm DM} \, y \right),
    \label{eq:Yeqdef}
\end{equation}
is the equilibrium yield, and
\begin{equation}
    Z(y) \equiv 
    \sqrt{\frac{\pi}{45}}M_{\rm Pl}
    \frac{\sqrt{g_*}\, \langle\sigma v\rangle}{m_{\rm DM}\,y^2}.
    \label{eq:Zdef}
\end{equation}
In Eqs.~\eqref{eq:Yeqdef} and \eqref{eq:Zdef}, the special function $K_2$ is the modified Bessel function of the second kind, $M_{\rm Pl}$ is the Planck mass, $\sqrt{g_*}=\frac{h_{\rm eff}}{\sqrt{g_{\rm eff}}}(1+\frac{T}{3 h_{\rm eff}}\frac{d h_{\rm eff}}{dT})$, and $h_{\rm eff}$ and $g_{\rm eff}$ are the effective entropy and energy degrees of freedom, respectively. 
Further, $\langle\sigma v\rangle$ is the total thermal-averaged 
DM annihilation cross-section, computed through the
Gondolo-Gelmini formula~\cite{Gondolo:1990dk}:
\begin{equation}
\langle\sigma v\rangle
\equiv
\frac{y}{8 m_{\rm DM}^4 K_2^2 \left(m_{\rm DM} \,y \right)}
\int_{4m_{\rm DM}^2}^\infty
\sigma(s-4m_{\rm DM}^2)
\sqrt{s}K_1 \left(\sqrt{s} \,y \right) \mathrm{d}s.
\end{equation}
Here, the special function $K_1$ is the 
modified Bessel function of the first kind, and
$\sigma$ is the total DM annihilation scattering cross-section.

Equation~\eqref{eq:beq_standard} can be solved numerically to find the present DM yield. All the results and parameter space presented in our study are based on solving the Boltzmann equation numerically, adapted to include the phase transition effects. However, to better grasp the main subject of this paper, it is useful to recall the semi-analytical \textit{freeze-out approximation}~\cite{Cline:2013gha,Steigman:2012nb}, which we employ only to demonstrate the subtleties of the analysis. In this approximation, one assumes a freeze-out instant $y_f \equiv 1/T_f$, before which
before which DM particles are in thermal equilibrium with the primordial bath (so $Y = \overline{Y}$), and after which they decouple, leading to their comoving number density freezing out (meaning $Y \gg \overline{Y}$). 
In this sense, it is convenient to reparameterize the yield as $Y_{\rm DM}\approx (1+\delta) \overline{Y}$, where $\delta$ is a monotonically increasing function of $y$.
In this approximation, the decoupling instant $y_f$ can be numerically obtained by iteration through the following expression~\cite{Cline:2013gha}
%
%
\begin{equation}
y_f = \log \left(e^y\, \frac{\delta_f (2 + \delta_f)}{ 1 + \delta_f} \frac{ Z \,\overline{Y}^2}{\left(- \frac{\mathrm{d} \overline{Y} }{\mathrm{d} y}\right)} \right)_{y=y_f},
\label{eq:x_f_interation}
\end{equation}
where $\delta_f$ is a $\mathcal{O}(1)$ number\footnote{For example,
{\tt micrOMEGAs}  \cite{Belanger:2001fz,Alguero:2023zol}
uses $\delta_f=1.5$.} 
that dictates the amount by which the yield at freeze-out $Y_f$ differs from the equilibrium yield at the freeze-out instant $\overline{Y}(y_f)$, through
$Y_f = (1+\delta_f)\overline{Y}(y_f)$.

After the freeze-out instant, the \textit{standard} approach is to neglect $\overline{Y}$ in Eq.~\eqref{eq:beq_standard} and integrate it explicitly for all $y > y_f$.
In turn, the present-day DM yield is given by~\cite{Cline:2013gha}:
\begin{equation}
	Y_{\rm DM}^{\rm today} \approx \frac{Y_f}{1+ Y_f A_{y_f \to \infty}},
    \label{eq:FO approx}
\end{equation}
where $A_{y_1 \to y_2}$ is a function that quantifies the evolution of the yield between two points $y_1$ and $y_2$, and is given by
\begin{equation}
	A_{y_1 \to y_2} = \int_{y_1}^{y_2} Z(y)\, \mathrm{d}y = 
	\sqrt{\frac{\pi}{45}} 
	M_{\mathrm{Pl}}
	\int_{y_1}^{y_2} \, \frac{\sqrt{g_*}\left< \sigma \mathrm{v}\right>}{m_{\mathrm{DM}}\,y^2}
	    \, \mathrm{d}y.
        \label{eq:A}
\end{equation}

Finally, one may calculate the present DM relic density through
\textcolor{black}{\begin{equation}
	\Omega h^2 = \frac{h^2 s^{\rm today}}{\rho_c}\,m_{\mathrm{DM}}\, Y_{\rm DM}^{\rm today}
	\approx 
	2.744\times 10^8 \, \left(\frac{m_{\mathrm{DM}}}{\mathrm{GeV}}\right)\, Y_{\rm DM}^{\rm today},
    \label{eq:relic}
\end{equation}
where $\rho_c \approx 1.053\,672\,h^2 \,\mathrm{GeV\,cm^{-3}}$ is the critical energy density of the Universe and $s^{\rm today} \approx 2891.2 \,\mathrm{cm^{-3}}$ is the present entropy density~\cite{ParticleDataGroup:2024cfk}.
}

\subsection{Relic density: the improved approach}
\label{subsec:improved}

As explained in the previous section, the \textit{standard} way to calculate the present DM relic density is to integrate the function $Z(y)$ from the $y$ value at decoupling ($y=y_f$) until the present epoch ($y \to \infty$). This is consistent with the idea that the Boltzmann equation is dominated by the DM depletion processes occurring at temperatures below freeze-out, and hence below the DM mass scale~\cite{Arcadi:2024ukq}.

However, when EWSB occurs \textit{after} the DM freeze-out, \textit{i.e.}, when $T_f > T_{\rm EWSB}$, special care must be taken in defining Eq.~\eqref{eq:A}. This is because the integrand of Eq.~\eqref{eq:A} depends on both $m_{\rm DM}$ and $\langle\sigma v\rangle$, whose values change across the EWSB transition. 
Note that EWSB is expected to occur at a temperature
of approximately $T_{\rm EWSB} \sim 160\,\mathrm{GeV}$~\cite{Laine:2015kra},\footnote{
\textcolor{black}{
As demonstrated in~\cite{Carena:2021onl}, the EWSB temperature 
may change depending on the size of the particle content
and on the size of the couplings their couplings to the SM Higgs boson.
However, in that same paper, the authors find that significant changes to
$T_{\rm EWSB}$ require an added number of singlet scalars of
order $\mathcal{O}(100)$. For that reason, we assume 
$T_{\rm EWSB} \approx 160$\,GeV throughout this paper.
}
}
while the freeze-out of a typical scalar DM candidate takes place around
$T_f \sim m_{\mathrm{DM}}/25$~\cite{Cline:2013gha}.
This implies that EWSB occurs after DM freeze-out when~\cite{Bhattacharya:2021rwh}
\begin{eqnarray}
 T_f > T_{\rm EWSB}
 \Leftrightarrow
 y_f < y_{\rm EWSB}
   &\implies& m_{\rm DM} > 160 \times 25 \ {\rm GeV} \sim 4\,{\rm TeV}\,.
  \label{eq:massbound}
\end{eqnarray}
One should therefore expect the \textit{standard} approach to break down for DM masses above 4\,TeV, motivating the need for an \textit{improved} approach to more accurately calculate the present relic density for heavy DM candidates.


To properly define the {\it improved} approach\,---\,which incorporates the effects of EWSB\,---\,and to facilitate the comparison with the {\it standard} approach\,---\,where DM evolution is computed entirely in the electroweak broken phase, as is conventionally done in the dark matter literature\,---\, we must introduce some notation.
Let $\left< \sigma \mathrm{v}\right>_{\rm bEWSB}$ be the thermal averaged cross-section before EWSB and $\left< \sigma \mathrm{v}\right>_{\rm aEWSB}$ that after EWSB. Further, let $\widetilde{m}_{\rm DM}$ denote the DM mass in the electroweak unbroken phase and $m_{\rm DM}$ denote the DM mass in the electroweak broken phase. Then, the terminology we will use throughout this paper is
\begin{enumerate}[(i)]
    \item $\left< \sigma v\right>_S$ : represents the total thermal averaged cross-section in the {\it standard} approach and is defined by the relation
    \begin{align}
        \left< \sigma v\right>_S(y)=\left< \sigma v\right>_{\rm aEWSB}(y),
    \end{align}
    throughout all cosmic history, \textit{i.e.}, over the full range of $y$.
    \item $\left< \sigma v\right>_I$ : denotes the total thermal averaged cross-section in the {\it improved} prescription, which differentiates contributions from channels operating before and after EWSB, leading to the relation
    \begin{align}
        \left< \sigma v\right>_I(y)=\left< \sigma v\right>_{\rm bEWSB}(y)\, 
        \Theta(y_{\rm{EWSB}}-y)+\left< \sigma v\right>_{\rm aEWSB}(y)\, 
        \Theta(y-y_{\rm{EWSB}}),
    \end{align}
    where $\Theta$ is the Heaviside theta function.\footnote{
    \textcolor{black}{
    A $\Theta(T - T_{\mathrm{EWSB}})$ switch corresponds to leading-order prescription to separate the unbroken and broken electroweak phases. Such phase-by-phase matching treatments are standard in early Universe setups where the Lagrangian changes across symmetry-breaking boundaries~\cite{Heeba:2018wtf,Baldes:2021aph}.}
    }
    \item $Y^\infty_S$ : denotes the present-day DM yield computed in the {\it standard} approach. Following Eqs.~\eqref{eq:FO approx} and \eqref{eq:A}, it is given as:
    \begin{equation}
        Y^\infty_S=\frac{Y^S_f}{1 + Y^S_f A^S_{{y^S_f} \to \infty}},
    \end{equation}
    where the DM yield computed at the \textit{standard} freeze-out instant $y_f^S$ is
    \begin{equation}
        Y_f^S = \left( 1 + \delta_f \right) \overline{Y}\left(y^S_f\right),
        \label{eq:YfS}
    \end{equation}
    and   
\begin{equation}
        A^S_{{y_1} \to y_2}=
		\sqrt{\frac{\pi}{45}} 
		\frac{M_{\mathrm{Pl}}}{m_{\rm DM}}
		\int_{y_1}^{y_2} \, 
		\frac{\sqrt{g_*}\left< \sigma \mathrm{v}\right>_{\rm aEWSB}}{y^2}\, \mathrm{d}y.
		\label{eq:A_S}
    \end{equation}
	\item $Y^\infty_I$ : represents the present-day yield computed in the {\it improved} approach, and is defined by
    \begin{equation}
        Y^\infty_I=\frac{Y^I_f}{1 + Y^I_f A^I_{{y^I_f} \to \infty}},
    \end{equation}
    where the DM yield computed at the \textit{improved} freeze-out instant $y_f^I$ is
    \begin{equation}
        Y_f^I = \left( 1 + \delta_f \right) \overline{Y}\left(y^I_f\right),
        \label{eq:YfI}
    \end{equation}
    and   
\begin{eqnarray}
	A^I_{{y_1} \to y_2} &=&
		\sqrt{\frac{\pi}{45}} 
		\frac{M_{\mathrm{Pl}}}{\widetilde{m}_{\rm DM}}
		\int_{y_1}^{y_2} \, 
		\frac{\sqrt{g_*}\left< \sigma v\right>_{\rm bEWSB}}{y^2}\,\Theta(y_{\rm{EWSB}}-y)\, \mathrm{d}y\nonumber\\[0.5em]
		&&+\ 
		\sqrt{\frac{\pi}{45}} 
		\frac{M_{\mathrm{Pl}}}{m_{\rm DM}} 
		\int_{y_1}^{y_2} \, 
		\frac{\sqrt{g_*}\left< \sigma v\right>_{\rm aEWSB}}{y^2}\,\Theta(y-y_{\rm{EWSB}})\, \mathrm{d}y.
    \label{eq:A_I}
	\end{eqnarray} 
	\item $\Omega h^2|_S$ : is the \textit{standard} DM relic density, defined as
	\textcolor{black}{\begin{equation}
		\Omega h^2|_S = 2.744\times 10^8 \, \left(\frac{m_{\mathrm{DM}}}{\mathrm{GeV}}\right)\, Y^\infty_S.
		\label{eq:relic_S}
	\end{equation}}
	\item $\Omega h^2|_I$ : is the \textit{improved} DM relic density, defined as
	\textcolor{black}{\begin{equation}
		\Omega h^2|_I = 2.744\times 10^8 \, \left(\frac{m_{\mathrm{DM}}}{\mathrm{GeV}}\right)\, Y^\infty_I.
		\label{eq:relic_I}
	\end{equation}}
\end{enumerate}

\subsection{Comparing approaches}
\label{subsec:comparison}

We use the relative difference between $\Omega h^2|_S$ and $\Omega h^2|_I$
to quantify the deviation between the {\it standard} and {\it improved} approaches.
Based on the definitions of the relic densities introduced in Section~\ref{subsec:improved}, we write the relative difference as
\begin{equation}
\delta \Omega h^2 
\equiv \frac{\Omega h^2 |_I - \Omega h^2 |_S}{\Omega h^2 |_I}
= Y_S^\infty\left( \alpha - \beta \right),
\label{eq:error}
\end{equation}
where we define
\begin{eqnarray}
\alpha &\equiv & \frac{1}{Y^S_f}-\frac{1}{Y^I_f},\label{eq:alpha}\\
\beta &\equiv & A^I_{{y^I_f} \to \infty}-A^S_{{y^S_f} \to \infty}\label{eq:beta1}.
\end{eqnarray}
The sign of $\delta \Omega h^2$
is determined by the sign of $\left(\alpha -\beta\right)$.
While $\alpha$ quantifies the difference in inverse of yields at freeze-out between the two approaches, $\beta$ is a function of the annihilation cross-sections, with the dependence following from Eqs.~\eqref{eq:A_S} and \eqref{eq:A_I}. Looking carefully at Eqs.~\eqref{eq:A_S} and \eqref{eq:A_I}, one may conclude that the quantity $A_{y_{\rm EWSB}\to\infty}$ is the same in both approaches. 
Quantifying the relative error in this manner highlights the
interplay between two significant factors:
\textit{the annihilation cross-sections} and \textit{the scalar masses},
which together contribute to the shift in the final relic densities between
the two approaches.

With these definitions established, in the following sections, we will analyze the evolution of the resulting dark matter relic densities and quantify the differences arising between the two approaches within a specific extension of the SM. To this end, we introduce next a simple model where the \textit{standard} and \textit{improved} approaches can have consequential differences. In any case, the possible influence of EWSB on the computation of DM relic density must be explored in the study of any freeze-out DM model, provided the mass of the DM candidate is above 4\,TeV.

%
\section{A case study}
\label{sec:casestudy}

\subsection{Model definition}
\label{subsec:definition}

We extend the SM scalar sector by two real scalar singlets $\phi$ and
$\chi$~\cite{Cline:2013gha,Steigman:2012nb,Basak:2021tnj}.
The new scalars are stabilized by a $Z_2 \times Z_2^\prime$ symmetry such that
\begin{align}
  &  \chi \overset{Z_2}{\longrightarrow}-\chi
  \ \ \ \ \ \ \ \ {\rm and} 
  \ \ \ \ \ \ \ \phi \overset{Z_2^\prime}{\longrightarrow}-\phi \, .
  \label{eq:qno}
\end{align}
Consequently, these fields couple not only to the SM Higgs doublet via portal couplings, but also to each other. Before EWSB, the most general renormalizable scalar potential reads:
\begin{equation}
\begin{split}
V_{\rm bEWSB} =&\  \mu_H^2 H^\dagger H +	\lambda_H \left( H^\dagger H  \right)^2 
+ \frac{1}{2}\mu_\phi^2 \phi^2 +	\frac{1}{4!} \lambda_\phi \phi^4
+ \frac{1}{2}\mu_\chi^2 \chi^2 +	\frac{1}{4!} \lambda_\chi \chi^4\\
&+ \frac{1}{2} \lambda_{H\phi} \left( H^\dagger H  \right) \phi^2 
+ \frac{1}{2} \lambda_{H\chi} \left( H^\dagger H  \right) \chi^2 
+ \frac{1}{4} \lambda_{\phi\chi} \phi^2 \chi^2.
\end{split} 
\end{equation}
At the minimum of the potential, the Hessian matrix is diagonal
and the squared masses of the scalar fields come out as
\begin{equation}
\widetilde{m}^2_H = \mu_H^2\, ,
\ \ \ \ \ \ \ \ 
\widetilde{m}^2_\phi = \mu_\phi^2\, ,
\ \ \ \ \ \ \ \ 
\widetilde{m}^2_\chi = \mu_\chi^2\, .
\label{eq:masses_before}
\end{equation}

After EWSB, both the Higgs doublet and $\phi$
acquire a vev, breaking the $SU(2)\times U(1)$ and
$Z_2^\prime$ symmetries, respectively.
In turn, the bilinear terms in $H$ and $\phi$ flip signs.
Explicitly, the vacuum and field configuration after EWSB 
in the unitary gauge is
\begin{align}
&&
H &\to 
\frac{1}{\sqrt{2}}\begin{pmatrix}
0 \\
v_H + h 
\end{pmatrix}, &
\phi &\to v_\phi + \phi\, ,
&
\chi &\to \chi\, .
&&
\end{align}
Notice that this vacuum configuration allows for
$h$ and $\phi$ to mix into the mass basis through
\begin{equation}
\begin{pmatrix}
h \\
\phi
\end{pmatrix}	
=
\begin{pmatrix}
c_\theta & -s_\theta\\
s_\theta & c_\theta
\end{pmatrix}
\begin{pmatrix}
h_1 \\
h_2
\end{pmatrix},
\label{rot_theta}
\end{equation}
where $c_\theta = \cos{\theta}$ and $s_\theta = \sin{\theta}$.
We study this mixing in detail in App.~\ref{app:Va},
and denote by $V_{\rm aEWSB}$ the potential after EWSB\footnote{If
$\lambda_{H \phi} = 0$, then there is no coupling between the two sectors
and $\theta=0$, \textit{c.f.} Eq.~\eqref{tan_theta}.}.
The squared masses of the scalar fields read
\begin{eqnarray}
m_{h_1}^2 &=& v_H^2 \lambda_H + \frac{1}{6} v_\phi^2 \lambda_\phi 
+ \frac{1}{6} 
\sqrt{\left( 6 v_H^2 \lambda_H - v_\phi^2 \lambda_\phi \right)^2
 + \left( 6 v_H v_\phi \lambda_{H\phi} \right)^2},
\label{11}
\\
m_{h_2}^2 &=& v_H^2 \lambda_H + \frac{1}{6} v_\phi^2 \lambda_\phi 
- \frac{1}{6} 
\sqrt{\left( 6 v_H^2 \lambda_H - v_\phi^2 \lambda_\phi \right)^2
 + \left( 6 v_H v_\phi \lambda_{H\phi} \right)^2},
\label{22}\\
 m_\chi^2 &=& 
 \mu_\chi^2 + \frac{1}{2}v_H^2 \lambda_{H\chi} + \frac{1}{2}v_\phi^2 \lambda_{\phi\chi} .
\label{33}
\end{eqnarray}
The scalars $h_1$ and $h_2$,
besides coupling to $\chi$,
have additional relevant interactions with
$W$ and $Z$ gauge bosons, and SM fermions through
\begin{equation}
\mathcal{L} \supset 
\left[ \frac{\left( c_\theta h_1 - s_\theta h_2 \right)}{v_H} + 
\frac{\left( c_\theta h_1 - s_\theta h_2 \right)^2}{2 v_H^2} \right] 
\left( 2 m_W^2 W^+ W^-  + m_Z^2 Z^2\right) 
- \left( c_\theta h_1 - s_\theta h_2 \right) \sum_f  \frac{m_f}{v_H} \bar f f.
\end{equation}

We chose as input parameters for this model
\begin{equation}
m_{h_1},\ 
m_{h_2},\ 
m_\chi,\ 
v_H,\ 
\sin\theta,\ 
\lambda_{\chi},\ 
\lambda_{H\phi},\ 
\lambda_{H\chi},\ 
\lambda_{\phi\chi},
\label{parameters_aEWSB}
\end{equation}
where we fix $v_H = 246$ GeV, and identify $h_1$ with the $125$\,GeV scalar found at the LHC~\cite{ATLAS:2012yve,CMS:2012qbp}. Note that the values of these parameters are chosen such that the scalar potential has a minimum both before and after EWSB. All other parameters can be written in terms of these; \textit{c.f.} App.~\ref{app:Va}.

\subsection{Dark sector evolution}
\label{subsec:evolution}

\begin{figure}[h]
    \centering
\includegraphics[width=0.15\linewidth]{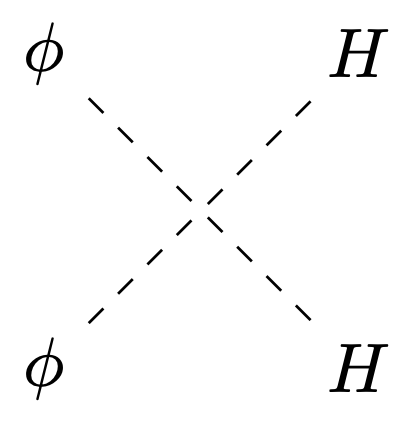}
\quad\quad\quad
\includegraphics[width=0.15\linewidth]{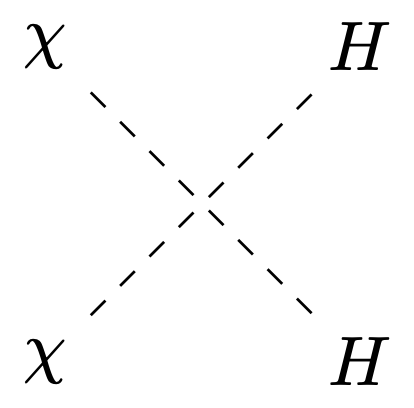}
\quad\quad\quad
\includegraphics[width=0.15\linewidth]{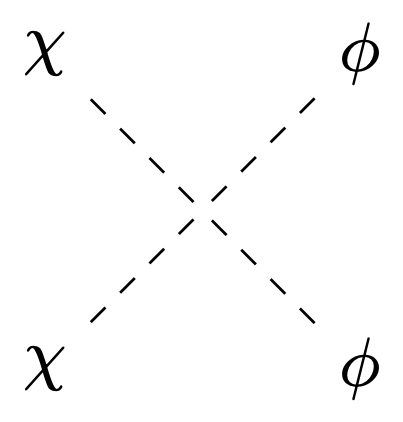}
\caption{Annihilation channels of scalar particles before EWSB}
    \label{fig:annihilation_bEWSB}
\end{figure}

In this model, when $T > T_{\mathrm{EW}}$, both $\phi$ and $\chi$ are
stable due to the unbroken $Z_2 \times Z_2^\prime$ symmetry.
This implies that there are two DM components,
each forming its own dark sector.
The possible annihilation channels are the quartic interactions 
$\phi \phi \to H H$,
$\chi \chi \to H H$, and 
$\chi \chi \to \phi \phi$,
shown in Fig.~\ref{fig:annihilation_bEWSB},
which are controlled by the quartic
couplings $\lambda_{H\phi}$, $\lambda_{H\chi}$ and $\lambda_{\phi\chi}$,
respectively.
Large values of these couplings keep the DM components in thermal equilibrium with the Higgs doublet $H$, which is itself in equilibrium with the SM bath because of sizable couplings with the SM gauge bosons and massless SM fermions.

If both components were to freeze out after EWSB, then, the first line of Eq.~\eqref{eq:A_I}
would vanish,
leading to
$A^I_{{y_1} \to y_2}=A^S_{{y_1} \to y_2}$ of Eq.~\eqref{eq:A_S},
and the two approaches would coincide.
This would only be achieved if the mass hierarchies were
$m_\chi,m_\phi \lesssim\ 4\, {\rm TeV}$,
due to the reason mentioned in Eq.~\eqref{eq:massbound}.
To simplify the analysis conservatively, in our study,
we impose $m_\phi \lesssim\, 1\ {\rm TeV}$,
so that $\phi$ remains in the thermal bath before EWSB.
As a result, $Y_\phi$ always traces $ \overline{Y}_\phi$ until EWSB occurs.
On the other hand,
we require $\widetilde{m}_\chi \gtrsim 4\ {\rm TeV}$,
such that $Y_\chi$ saturates before EWSB.
Thus, the treatment of $\chi$ in the \textit{standard} and \textit{improved} approaches will
have a noticeable physical consequence.

\begin{figure}[!h]
\centering
\raisebox{-.5\height}{\includegraphics[width=0.18\linewidth]{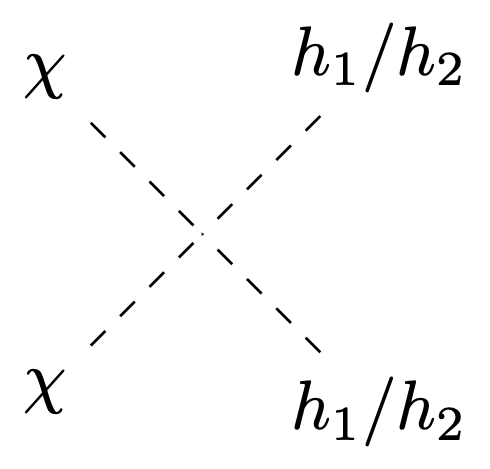}}
\quad\quad
\raisebox{-.5\height}{\includegraphics[width=0.21\linewidth]{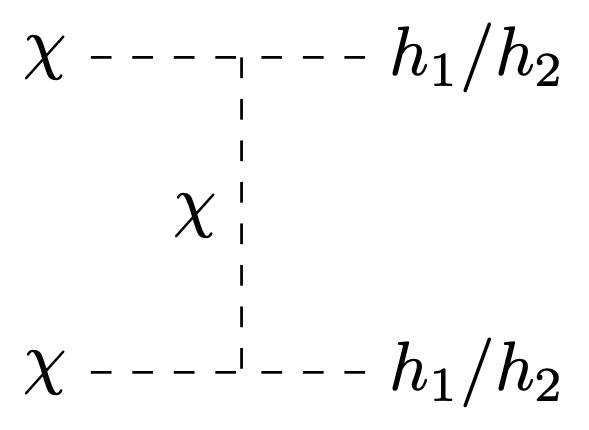}}
\quad\quad\quad
\raisebox{-.5\height}{\includegraphics[width=0.21\linewidth]{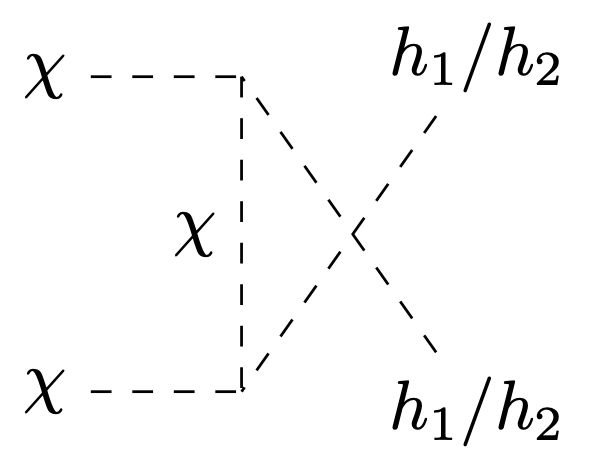}}
\quad\quad
\raisebox{-.5\height}{\includegraphics[width=0.21\linewidth]{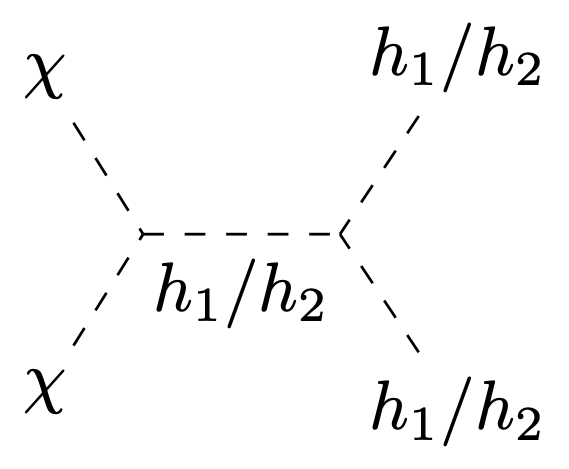}}\\
\raisebox{-.5\height}{\includegraphics[width=0.18\linewidth]{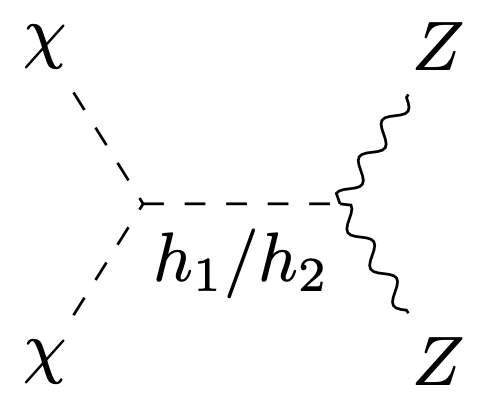}}
\quad\quad\quad
\raisebox{-.5\height}{\includegraphics[width=0.18\linewidth]{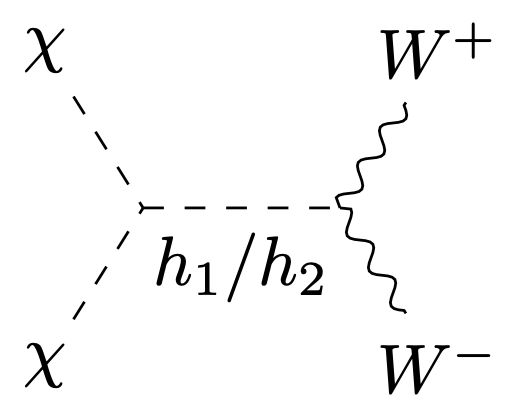}}
\quad\quad\quad
\raisebox{-.5\height}{\includegraphics[width=0.18\linewidth]{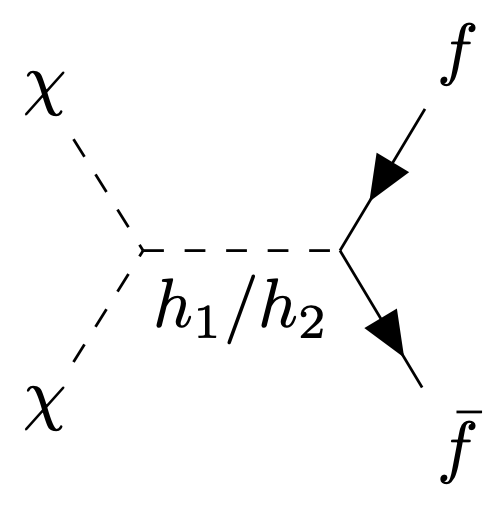}}
    \caption{Dark matter annihilation channels after EWSB.}
    \label{fig:annASB}
\end{figure}

When $T< T_{\mathrm{EW}}$, EWSB kicks in 
and the scalars $H$ and $\phi$ acquire their respective 
vevs, the $Z_2^\prime$ symmetry is explicitly broken, 
no longer rendering $\phi$ as a DM component.
In this electroweak broken phase,
the mixing with the CP-even 
component of the Higgs doublet allows $h_2$
to decay into SM particles.
These decay channels cause $Y_{h_2}$ to eventually drop
to very low values as the temperature of the Universe begins to fall.
This implies that, after EWSB,
$\chi$ becomes the only DM component, still stabilized through the discrete
symmetry assigned in Eq.~\eqref{eq:qno}. 
Also, because of the mixing between scalars in this regime,
several additional channels open up as shown in Fig.~\ref{fig:annASB},
which feature the pair annihilation of $\chi$ into SM fermions and gauge
bosons through $s$-channel processes mediated by $h_1$ and $h_2$. 

As introduced in Sec.~\ref{sec:relic}, the {\it standard} approach assumes
DM to evolve entirely in the electroweak broken phase, while the {\it improved}
approach distinguishes between the two phases\,---\,before EWSB and after
EWSB\,---\,incorporating the full thermal history.
It is worth reiterating here that the physical mass of DM is $m_\chi$ for the
{\it standard} approach throughout the thermal history of the Universe,
whereas the {\it improved} formalism considers DM mass as $\widetilde{m}_\chi$
until EWSB and $m_\chi$ afterward. 
Therefore, differences in relic density between the two approaches can arise 
either from changes in the DM annihilation cross section 
or from changes in the scalar masses across the two phases.

%
\section{Results and observations}
\label{sec:results}

\subsection{Numerical scan}
\label{subsec:scan}

To better understand the extent to which the DM relic density differs between the \textit{standard} and \textit{improved} approaches, we performed a random scan of $10^6$ benchmark points to probe the region of parameter space bounded by:
\begin{align}
m_{\chi} &\in \left[ 1,100 \right]\,\mathrm{TeV}, &
m_{h_2} &\in \left[ 250,1000 \right]\,\mathrm{GeV}, &\nonumber\\
\sin\theta  &\in \left[ -0.3 , 0.3 \right],&
\lambda_{\chi},
\lambda_{H\phi},
\lambda_{H\chi},
\lambda_{\phi\chi} &\in \pm\left[ 10^{-3}, 4\pi \right].
\end{align}
As explained, the upper bound on the mass of $h_2$ is chosen to ensure it remains in thermal equilibrium with the SM bath until EWSB occurs.
The range considered for the mixing angle $\theta$ is consistent with the latest experimental bounds on heavy Higgs-like scalar mixing~\cite{ATLAS:2021pkb}.
Regarding quartic couplings, we scanned over their 
typical freeze-out values, ensuring they respect perturbativity
and the boundedness from below conditions of App.~\ref{app:BFB}.
We confirmed that, for values of $\lambda$'s below $10^{-3}$, there are no new features, but occasional numerical instabilities may occur.

To calculate the relic density of $\chi$ in the \textit{standard} approach,
we implemented, as usual, the model described in Sec.~\ref{sec:casestudy}
in the electroweak broken phase. We used {\tt micrOMEGAs 6.0.5}~\cite{Alguero:2023zol}
to numerically integrate the Boltzmann Equation and track the evolution
of the yield of $\chi$ from $T=10^5$\,GeV to $T=10^{-4}$\,GeV.
The \textit{standard} relic density was then calculated using Eq.~\eqref{eq:relic_S}.
On the other hand, the calculation of the relic density in the \textit{improved} approach was performed in two steps. In the first step, we implemented the model before EWSB and used {\tt micrOMEGAs} to integrate the Boltzmann Equation from $T=10^5$\,GeV to $T=T_{\rm EWSB}$.
In the second step, we used the model after EWSB and the previously calculated yield at $T=T_{\rm EWSB}$ as inputs to the {\tt darkOmegaNTR()} function of {\tt micrOMEGAs}, allowing us to integrate the Boltzmann Equation for $\chi$ from $T=T_{\rm EWSB}$ to $T=10^{-4}$\,GeV. Finally, the \textit{improved} relic density was then calculated using Eq.~\eqref{eq:relic_I}.

\begin{figure}
\centering
\includegraphics[width=0.6\textwidth]{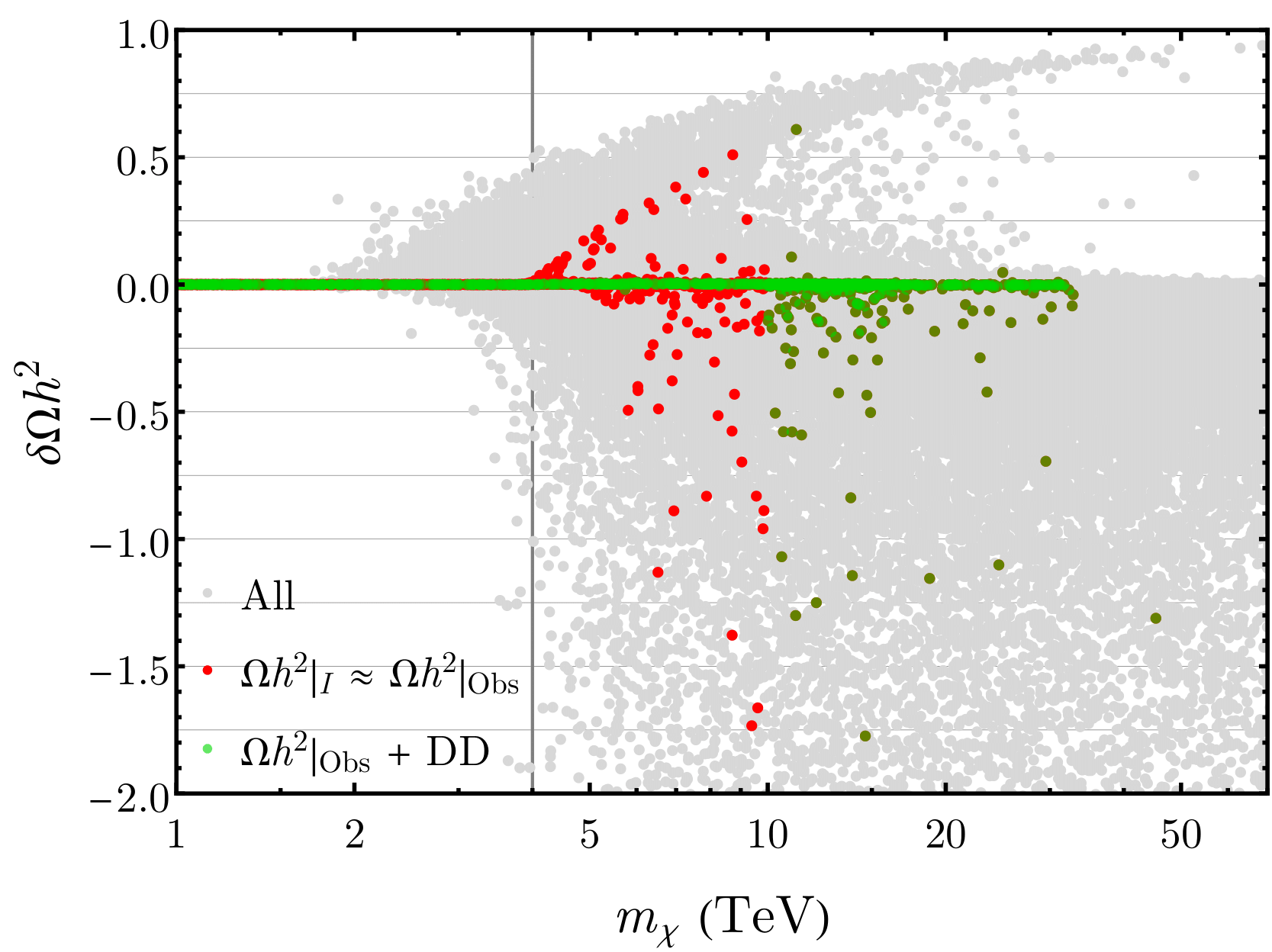}
\caption{Relative deviation between approaches ($\delta\Omega h^2$) against $m_\chi$. Red points have $\Omega h^2 |_I \approx 0.120$.
Green points have $\Omega h^2 |_I \approx 0.120$ and are allowed by Direct Detection constraints.}
\label{fig:observed_relic_ex2}
\end{figure}

In Fig.~\ref{fig:observed_relic_ex2}, we scan over the parameter space and plot, in gray, the relative difference ($\delta \Omega h^2$) between the {\it standard} and {\it improved} approaches as a function of the dark matter mass, over the range $1–100$\,TeV.
Here, the gray vertical line marks the inverse of the EWSB temperature.
Points with relic density computed in the {\it improved} approach and lying within the $3\sigma$ range of the PLANCK measurement~\cite{Planck:2018vyg} are shown in red.
As expected, $\delta \Omega h^2$ begins to deviate noticeably from zero as $m_\chi$ approaches 4 TeV\,---\,the approximate lower bound on the dark matter mass required for it to decouple from the SM bath before electroweak symmetry breaking, as discussed in Sec.~\ref{sec:relic}.
A subset of the red points, marked with green, satisfy the additional requirement that the direct detection (DD) cross-section lies below the current experimental limits. 
\textcolor{black}{We applied the latest exclusion limits on the spin-independent DM–nucleon scattering cross-section from the 2024 LUX-ZEPLIN (LZ) collaboration~\cite{LZ:2024zvo} across the entire mass range considered.
However, these limits are sufficiently constraining only for $m_\chi \lesssim 10,\mathrm{TeV}$. 
For the mass range $m_\chi \gtrsim 10,\mathrm{TeV}$, the present experimental sensitivity is significantly weaker and does not meaningfully restrict the parameter space~\cite{Bhoonah:2020fys,LZ:2024psa}. 
An explicit formula for the spin-independent scattering cross-section is provided in App.~\ref{app:direct}.}

\textcolor{black}{In Table.~\ref{tab:BPs}, we show two red benchmark points with $m_\chi < 10\,{\rm TeV}$ with non-zero $\delta\Omega h^2$ that have DD cross-sections below the 2024 LZ bounds~\cite{LZ:2024zvo}.
We found that the green points tend to favour small values of $\delta\Omega h^2$, of order $\mathcal{O}(1\%)$, reflecting the fact that in this simplified model, both the relic density and the DD cross-section are controlled by the scalar quartic couplings $\lambda_{H\chi}$ or $\lambda_{\phi\chi}$.
These couplings need to be large to achieve the observed relic density, but small enough to pass DD constraints.
However, we expect these exclusions to be less restrictive in more complete models, where the relic abundance and DD channels are not necessarily correlated.}

\textcolor{black}{In contrast, Table~\ref{tab:BPs2} presents two red benchmark points with $m_\chi > 10,\mathrm{TeV}$, where current DD limits are significantly less constraining. 
In this heavy-mass regime, the difference between the standard and improved approaches can reach $\mathcal{O}(100\%)$.}

\begin{table}
\centering
\renewcommand{\arraystretch}{1.5}
\setlength{\tabcolsep}{4pt}
\centering
\small
\begin{tabular}{|c|c|c|c|c|c|c|c|c|c|c|}
\hline
\multicolumn{2}{|c|}{\textbf{Masses}}&
\multicolumn{4}{|c|}{\textbf{Mixing and Couplings}}&
\multicolumn{3}{|c|}{\textbf{Relic Abundance}} \\
\hline\hline
$m_\chi$ (TeV) &
$m_\phi$ (GeV)&
$\sin\theta$ &
$\lambda_{H\phi}$ &
$\lambda_{H\chi}$ &
$\lambda_{\phi\chi}$ &
$\Omega h^2|_I$ &
$\Omega h^2|_S$ &
$|\delta\Omega h^2|\, (\%)$ \\
\hline
5.737&
984&
0.001 &
0.003 &
0.003 &
3.635 &
0.122721 &
0.125122 &
1.95 \\
\hline
8.281&
467&
0.036 &
$-0.095$ &
$-0.248$ &
5.103 &
0.118672 &
0.119259 &
0.49 \\
\hline
\end{tabular}
\caption{Benchmark points with $m_\chi \lesssim\, 10\ {\rm TeV} $ with finite $\delta\Omega h^2$ and direct detection cross-sections below LZ limits.}
\label{tab:BPs}
\end{table}

\begin{table}
\centering
\renewcommand{\arraystretch}{1.5}
\setlength{\tabcolsep}{4pt}
\centering
\small
\begin{tabular}{|c|c|c|c|c|c|c|c|c|c|c|}
\hline
\multicolumn{2}{|c|}{\textbf{Masses}}&
\multicolumn{4}{|c|}{\textbf{Mixing and Couplings}}&
\multicolumn{3}{|c|}{\textbf{Relic Abundance}} \\
\hline\hline
$m_\chi$ (TeV) &
$m_\phi$ (GeV)&
$\sin\theta$ &
$\lambda_{H\phi}$ &
$\lambda_{H\chi}$ &
$\lambda_{\phi\chi}$ &
$\Omega h^2|_I$ &
$\Omega h^2|_S$ &
$|\delta\Omega h^2|\, (\%)$ \\
\hline
11.146&
785&
$-0.055$ &
0.048 &
0.761 &
7.714 &
0.122207 &
0.128158 &
130.07 \\
\hline
11.175&
356&
0.207 &
0.003 &
6.608 &
0.291 &
0.119863 &
0.046840 &
60.92 \\
\hline
\end{tabular}
\caption{Benchmark points with $m_\chi >\, 10\ {\rm TeV} $ with finite $\delta\Omega h^2$ and direct detection cross-sections below experimental limits.}
\label{tab:BPs2}
\end{table}

\begin{figure}[h!]
\centering
\raisebox{-.55\height}{\includegraphics[width=0.6\textwidth]{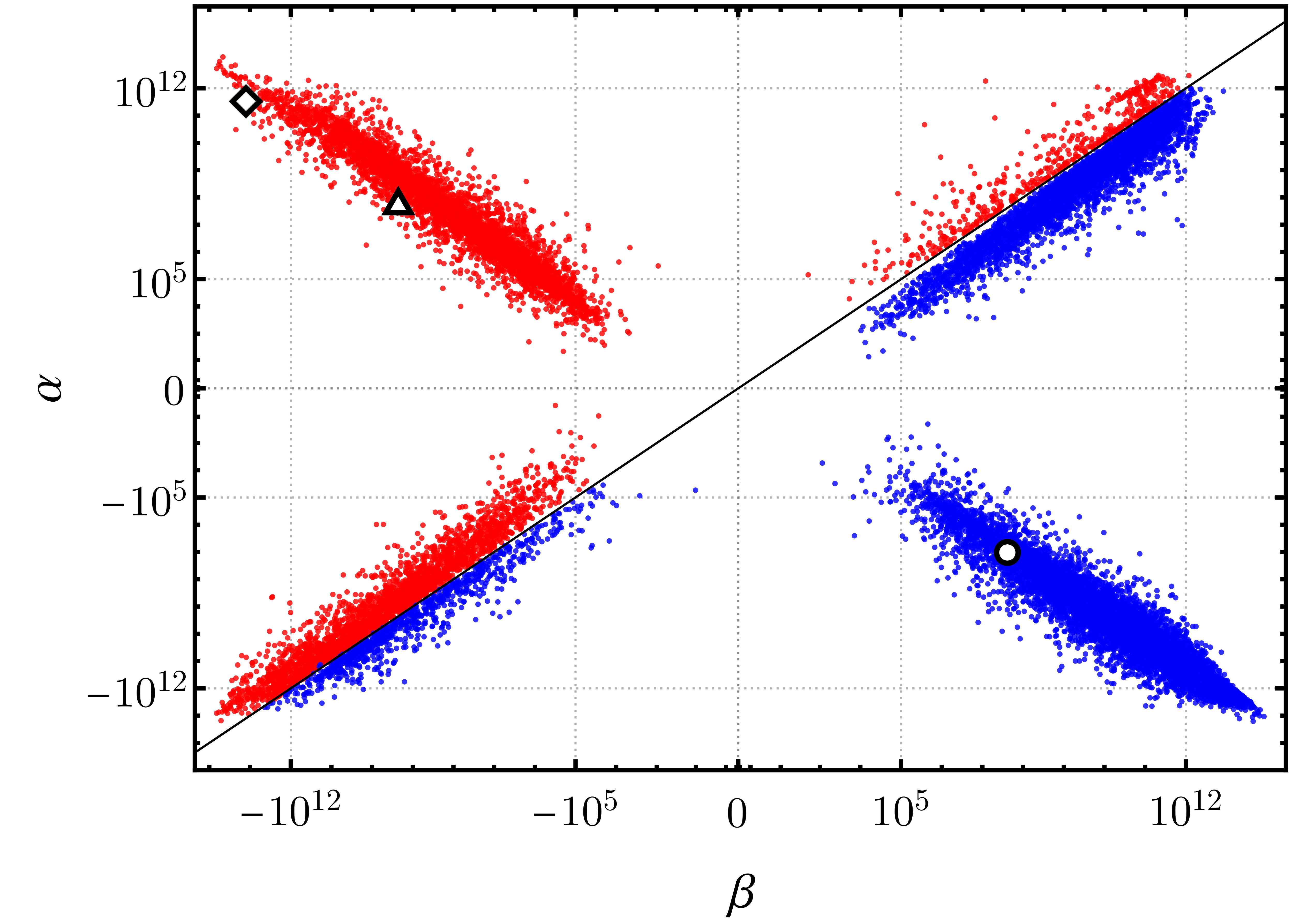}}
\quad
\raisebox{-.5\height}{\includegraphics[width=0.13\textwidth]{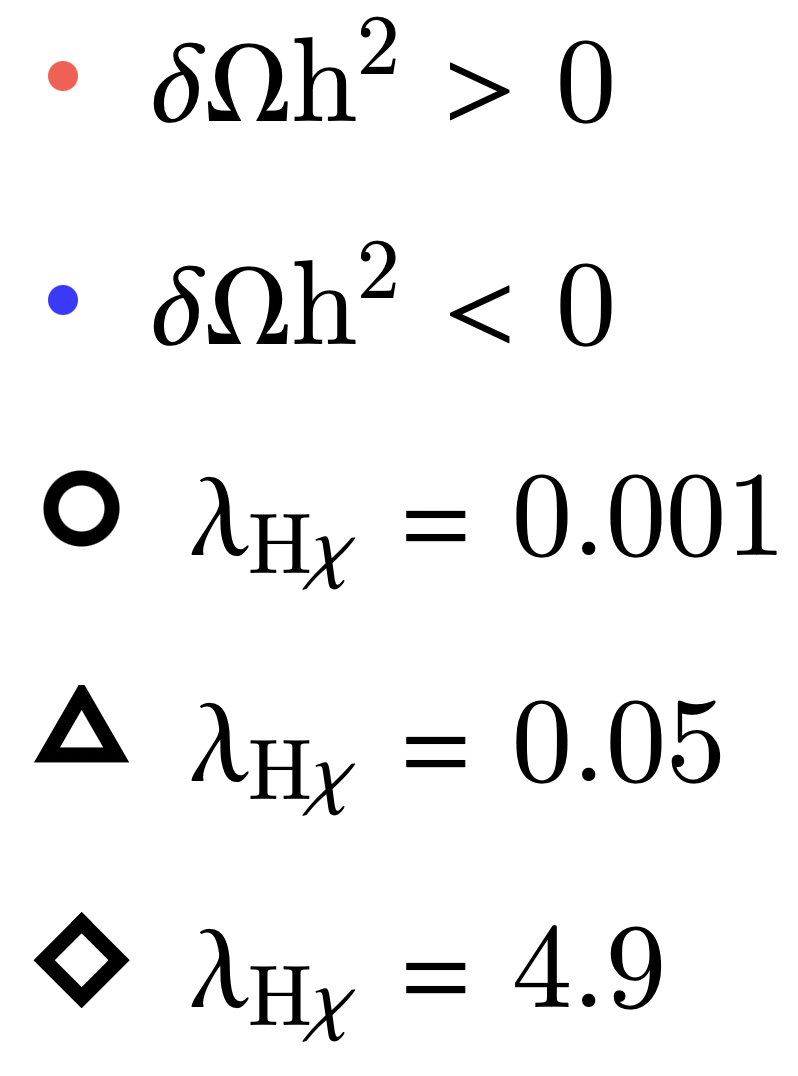}}
\caption{
Sign of $\delta \Omega h^2$ plotted in the $\alpha-\beta$ plane.}
\label{fig:sign_d_OMEGA}
\end{figure}

Notice that the sign of $\delta \Omega h^2$ may either be positive or negative,
as a consequence of the \textit{improved} relic density lying
above or below the \textit{standard} relic density, respectively.
The sign of the relative deviation between approaches
can be predicted using Eq.~\eqref{eq:error}.
To confirm the extent to which the numerical evaluation performed in 
{\tt micrOMEGAs} agrees with the \textit{improved} semi-analytical formalism
introduced in Sec.~\ref{subsec:improved},
we compute $\alpha$ and $\beta$ using to Eqs~\eqref{eq:alpha} 
and \eqref{eq:beta1} and use these values to position each point scanned 
on the $\alpha-\beta$ plane, as shown in Fig.~\ref{fig:sign_d_OMEGA}.
The color assigned to each point is based on the output from {\tt micrOMEGAs}.
We use red to indicate points with $\delta \Omega h^2 > 0$
and blue for points with $\delta \Omega h^2 < 0$.
The black solid line delineates the boundary $\alpha = \beta$.
The excellent agreement with the numerical results 
obtained from {\tt micrOMEGAs} demonstrates 
that the \textit{improved} approach reliably predicts the sign of the deviation.

\begin{figure}[h]
\centering
\includegraphics[width=0.47\textwidth]{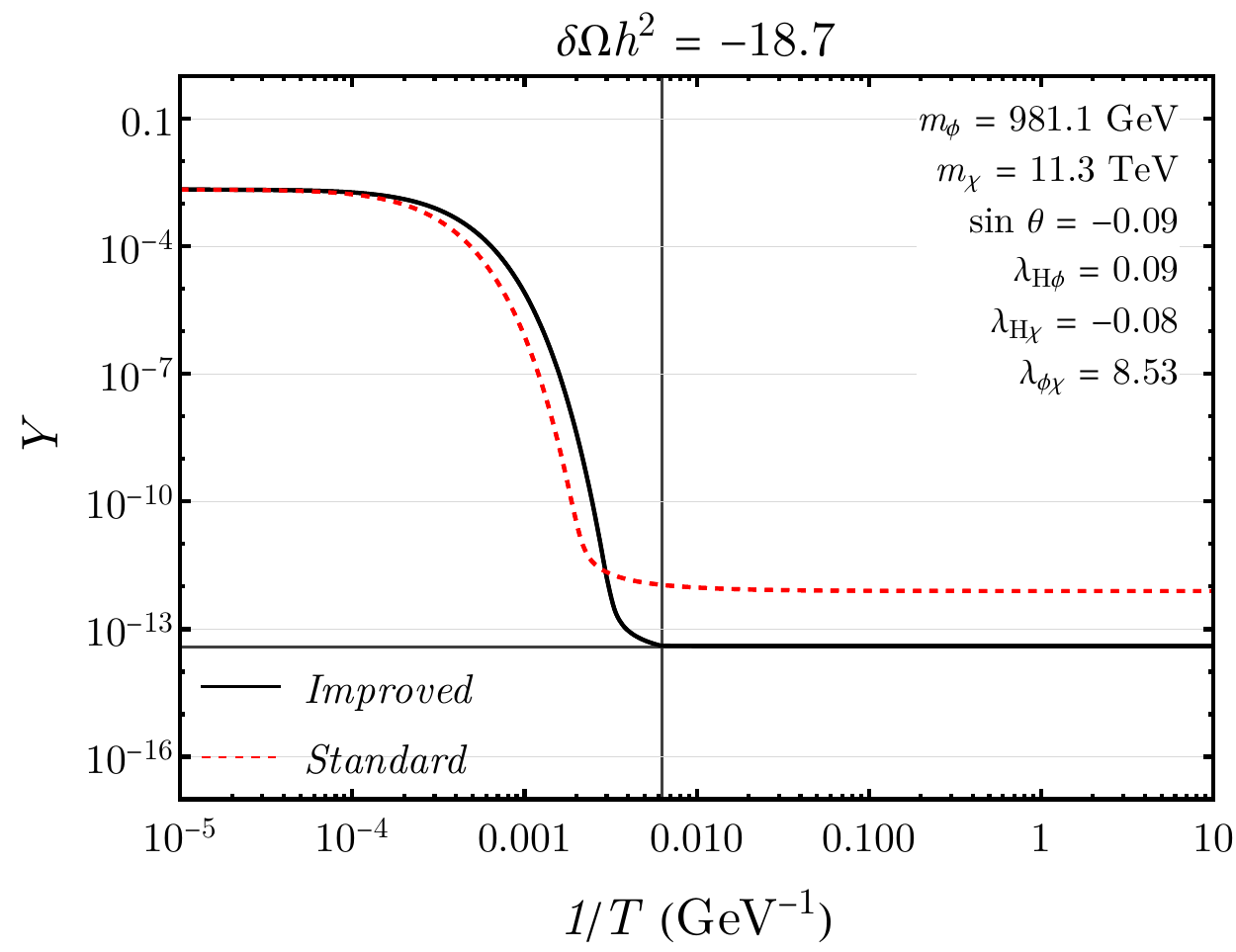}
\quad\quad
\includegraphics[width=0.47\textwidth]{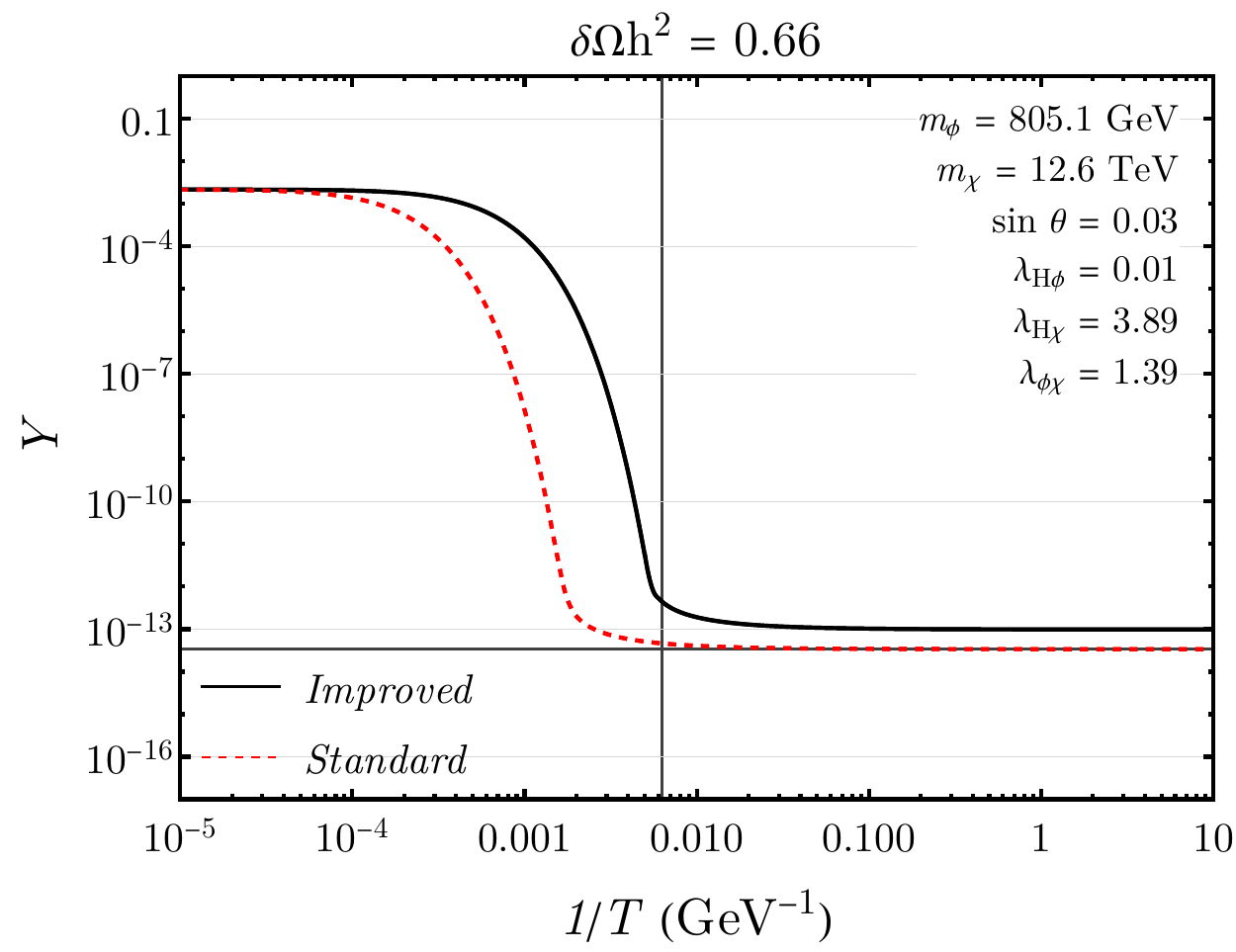}
\caption{
Left: A point discarded by the {\it standard} formalism but allowed in the {\it improved} approach.
Right: A point allowed in the \textit{standard} approach, but the {\it improved} approach forbids it.}
\label{fig:observed_relic}
\end{figure}

To further highlight the need for a re-evaluation of DM relic density computations in TeV-scale dark sectors, in Fig.~\ref{fig:observed_relic} we present the thermal history of $Y_\chi$ for two benchmark points. 
In these plots, the gray vertical line indicates the inverse of the EWSB temperature, and the gray horizontal line marks the yield corresponding to the observed relic density.
The left panel illustrates a case where the final yield in the {\it standard} approach is overabundant and therefore excluded, but matches the observed relic density in the {\it improved} approach, thereby reopening otherwise disfavored regions of parameter space. The right panel features the opposite scenario, where a relic abundance deemed acceptable in the {\it standard} formalism turns out to be over-abundant in the {\it improved} approach, thereby constraining the parameter space further. 

\subsection{Picking a benchmark point}
\label{subsec:bmp}

To better understand how changes in the dark matter 
annihilation cross-section and variations in scalar 
masses across the electroweak phase transition affect 
the relic density calculation, we consider a new benchmark scenario
in which both effects are governed by a single parameter, 
$\lambda_{H\chi}$. We fix the remaining parameters as
\begin{equation}
m_{\chi} =  11.9\,\textrm{TeV}\, ,
\ \ 
m_{h_2} = 895.1 \,\textrm{GeV}\, ,
\ \ 
\sin{\theta} =  0.11\, ,
\ \ 
\lambda_\chi = 1.000\, ,
\ \ 
\lambda_{H\phi} = 0.002\, ,
\ \ 
\lambda_{\phi\chi} = 0.007\,.
\end{equation}

\begin{figure}[h]
	\centering
	\includegraphics[width=0.32\linewidth]{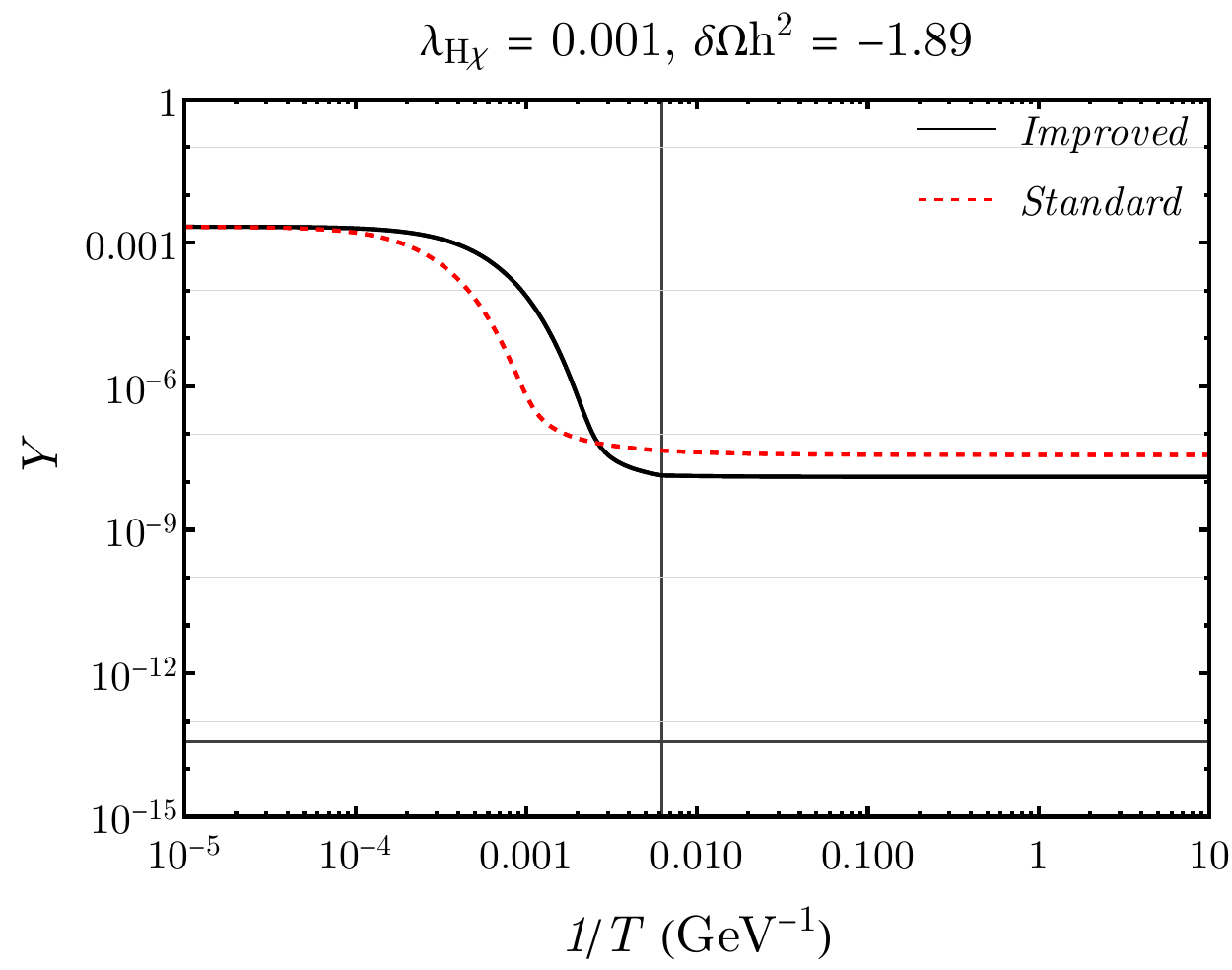}
	\includegraphics[width=0.32\linewidth]{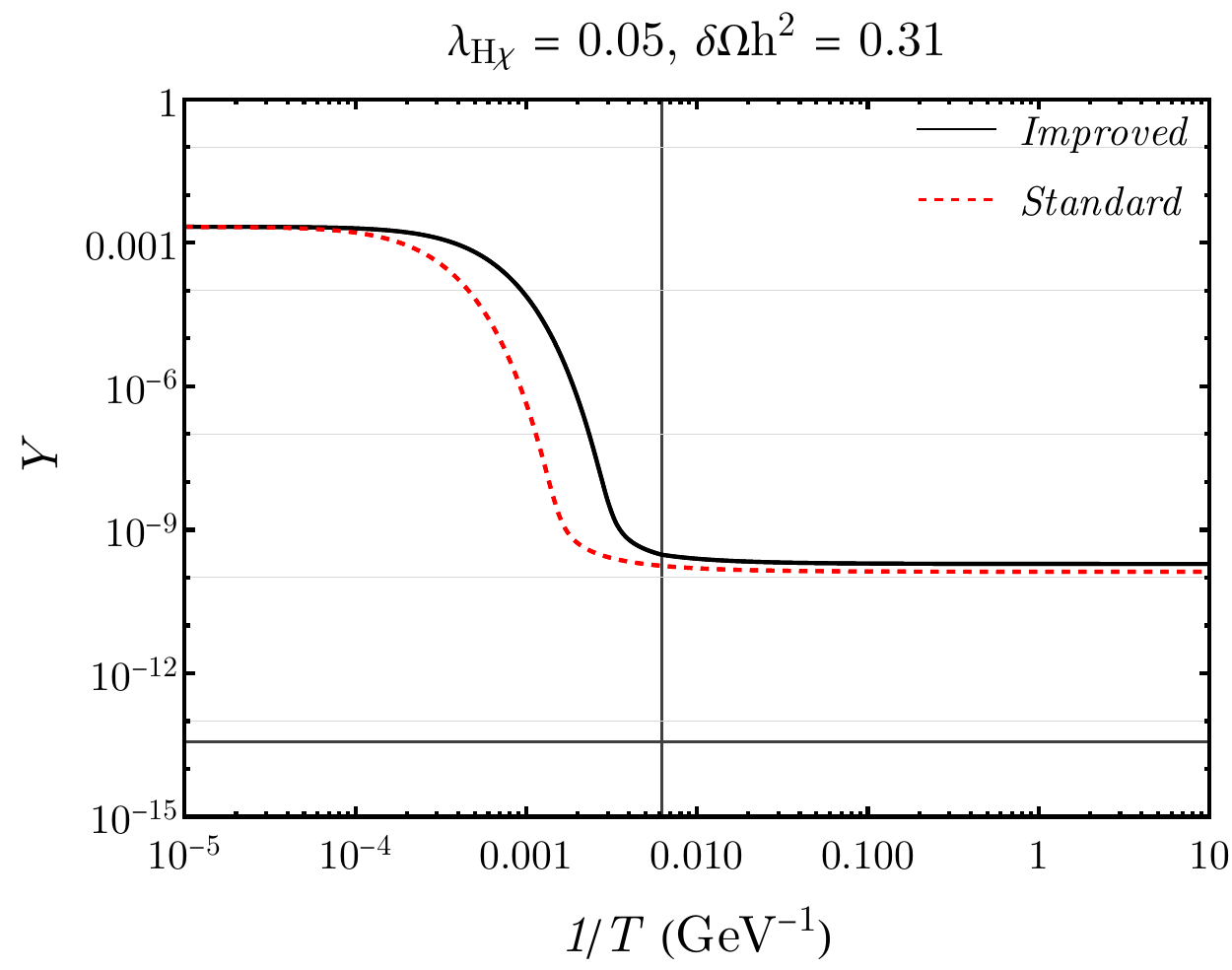}
	\includegraphics[width=0.32\linewidth]{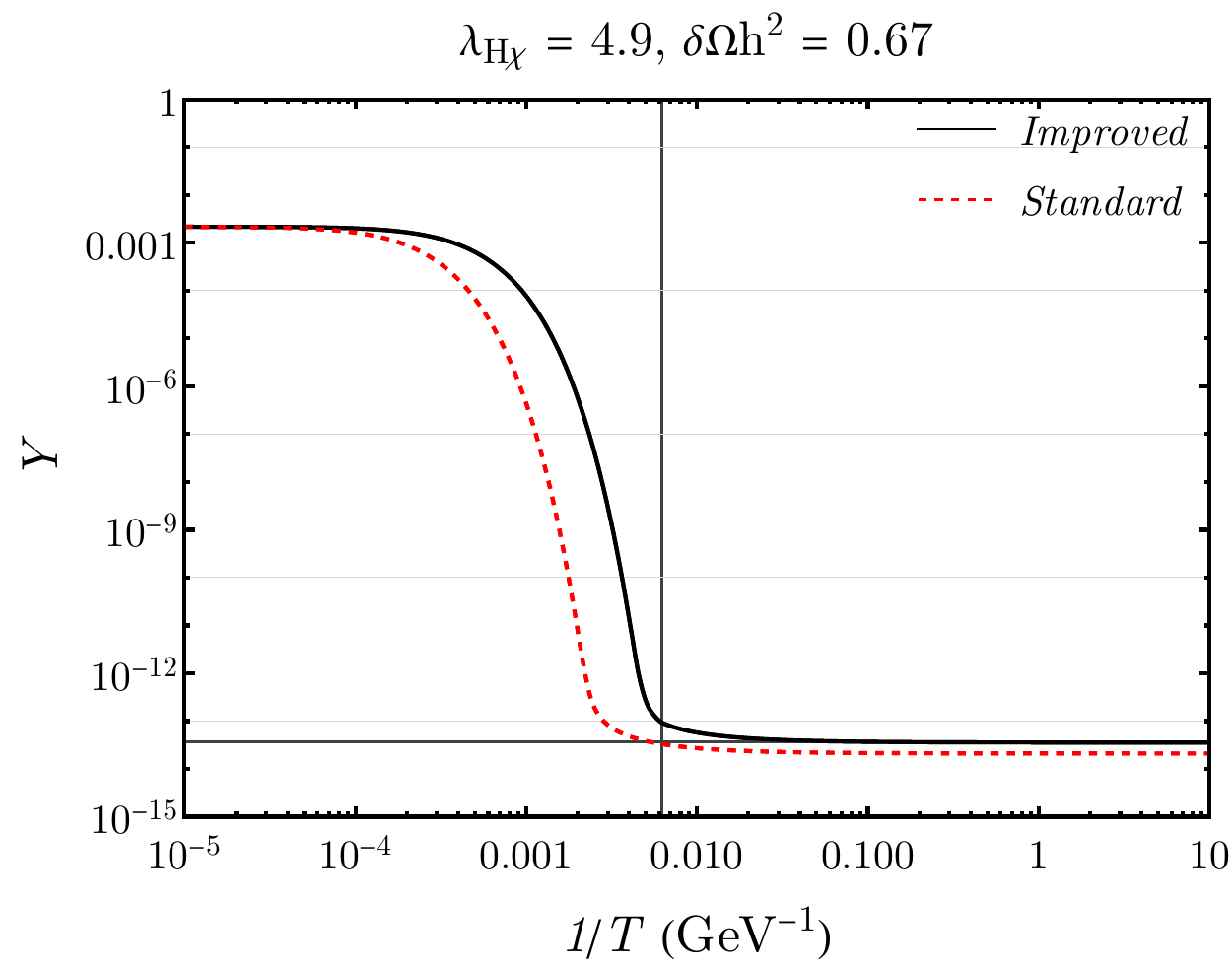}\\
	\includegraphics[width=0.32\linewidth]{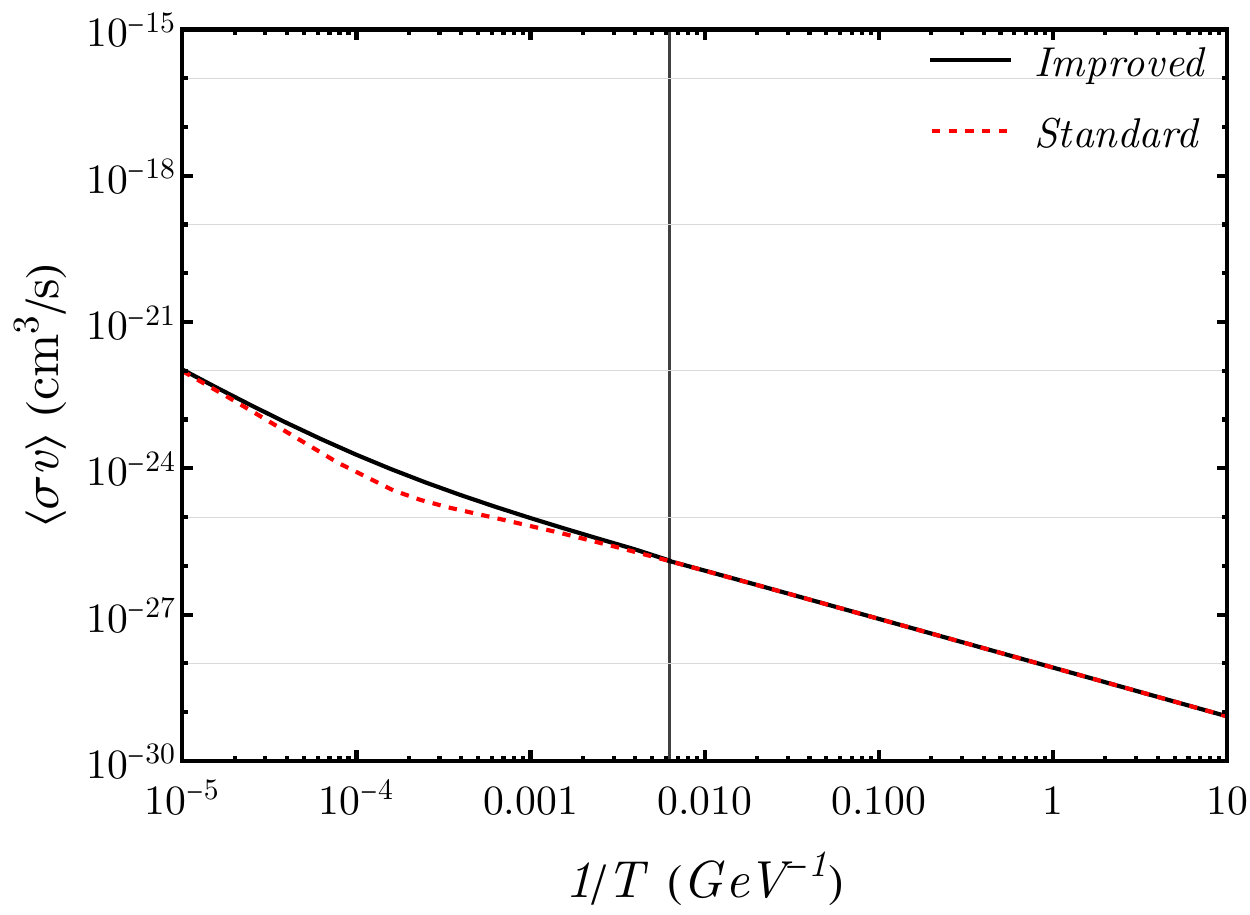}
	\includegraphics[width=0.32\linewidth]{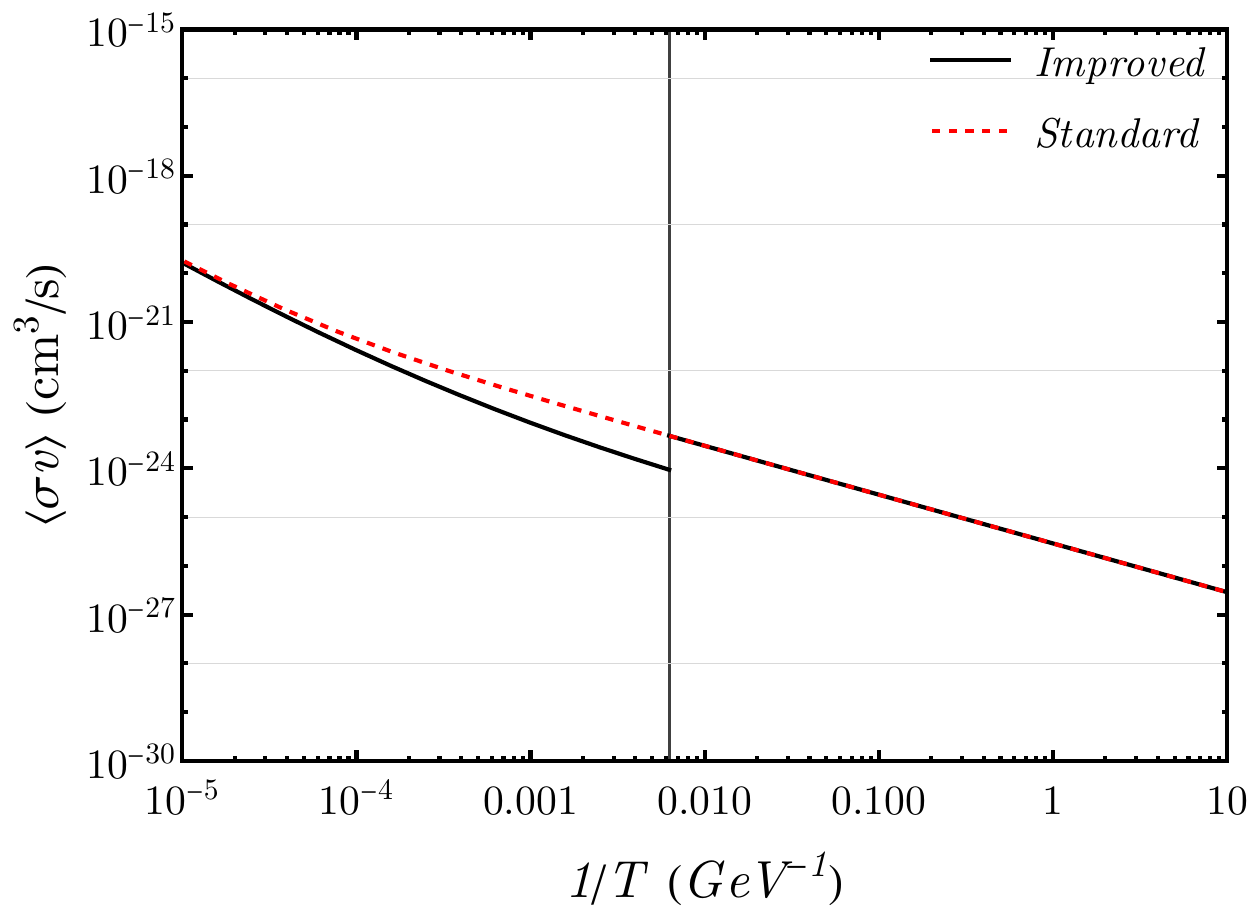}
	\includegraphics[width=0.32\linewidth]{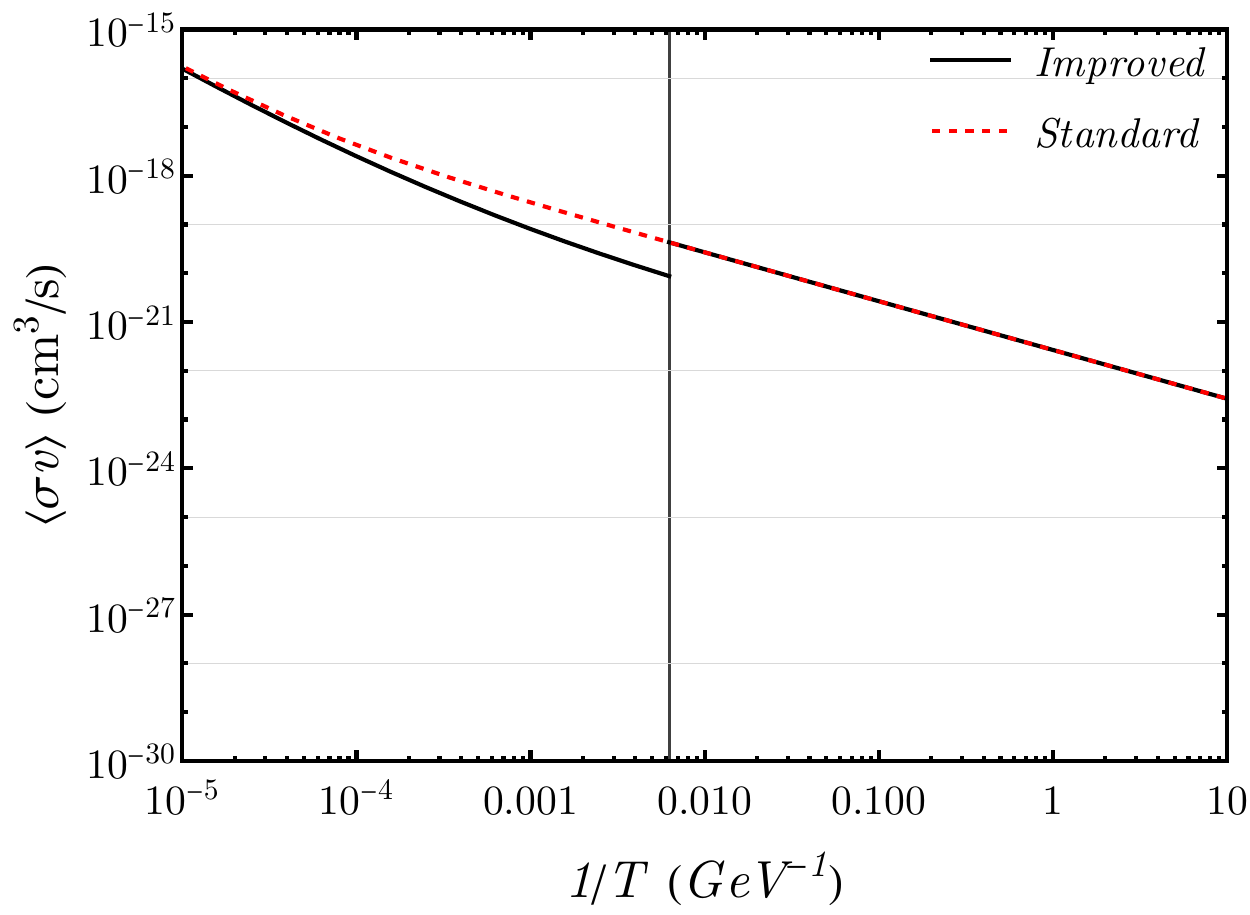}
	\caption{On the top (bottom) row of plots, the yield of $\chi$ (total thermal averaged cross-section) is plotted as a function of inverse temperature with $\lambda_{H\chi}$ increasing from left to right. The gray vertical line marks the inverse of $T_{\rm EWSB}$, and the grey horizontal line depicts the yield value corresponding to $\Omega h^2=0.12$.}
	\label{fig:vary_lamHC}
\end{figure}
In Fig.~\ref{fig:vary_lamHC}, we compare the DM yields predicted in the \textit{standard} and \textit{improved} approaches, as $\lambda_{H\chi}$ is increased.
The top row of plots shows the evolution of the yield of $\chi$ as a function of $1/T$, with $\lambda_{H\chi}$ increasing from left to right.
The black solid line represents the yield computed in the {\it improved} approach, the red dashed line represents the yield in the {\it standard} approach, the gray vertical line indicates the inverse of the EWSB temperature, and the gray horizontal line marks the yield corresponding to the observed relic density.
For each value of $\lambda_{H\chi}$, we trace in the bottom row of plots the corresponding temperature evolution of $\langle \sigma v \rangle_I$ with a black solid line, and that of $\langle \sigma v \rangle_S$ with a red dashed line.
It is interesting to note that $\chi$ always decouples from the SM bath earlier than in the \textit{improved} approach, implying the relation $y_f^S \lesssim y_f^I$.
Yet, regardless of the value of $\lambda_{H\chi}$, we always observe $\chi$ decoupling from the SM bath when $T>T_{\rm EWSB}$, as expected from the DM mass requirement in Eq.~\eqref{eq:massbound}.

In the first column of Fig.~\ref{fig:vary_lamHC}, corresponding to $\lambda_{H\chi}=0.001$, we observe that $\chi$ freezes out with a smaller yield in the \textit{improved} approach compared to the \textit{standard} one, leading to an overall negative $\delta \Omega h^2$.
This behavior can be understood as follows. 
At freeze-out, the \textit{standard} and \textit{improved} yields satisfy $Y_f^S > Y_f^I$, which implies a negative value for $\alpha$.
On the other hand, before EWSB, we notice that $\langle\sigma v \rangle_S \lesssim \langle\sigma v \rangle_I$, which implies that $A^S_{{y^I_f} \to \infty} < A^I_{{y^I_f} \to \infty}$ and results in a positive $\beta$.
For clarity, the particular values of $\alpha$ and $\beta$ for this benchmark point are marked by the white circular symbol in Fig.~\ref{fig:sign_d_OMEGA}.
A similar conclusion can be drawn for the other two benchmark points highlighted in Fig.~\ref{fig:sign_d_OMEGA}, which correspond to the $\lambda_{H\chi}$ values used in the middle and right-most columns of Fig.~\ref{fig:vary_lamHC}.
These two points, denoted by a triangle and a diamond in Fig.~\ref{fig:sign_d_OMEGA}, exhibit positive $\alpha$.
This arises from the fact that, before EWSB, the total thermal averaged cross-sections follow $\langle \sigma v \rangle_S > \langle \sigma v \rangle_I$, leading to a faster depletion of DM particles in the {\it standard} approach, and resulting in $Y_f^S < Y_f^I$. 
Due to the aforementioned tension between relevant parameters, $\beta$ remains negative in both cases. 
As a result, $\delta \Omega h^2$ turns out to be positive, in agreement with the values of $\delta \Omega h^2$ quoted on each row, and with the position of the solid black line ({\it improved} yield) lying above the red dashed one ({\it standard} yield).
Note that for $\lambda_{H\chi} = 4.9$, the \textit{improved} relic density
matches the observed value $\Omega h^2|_{\rm Obs} = 0.12$ at the $3\sigma$ level,
whereas the relic density computed in the \textit{standard} approach does not.
In fact, the \textit{standard} relic density deviates from $\Omega h^2|_{\rm Obs}$ by $42\sigma$.

\begin{figure}[h]
	\centering
	\includegraphics[width=0.32\linewidth]{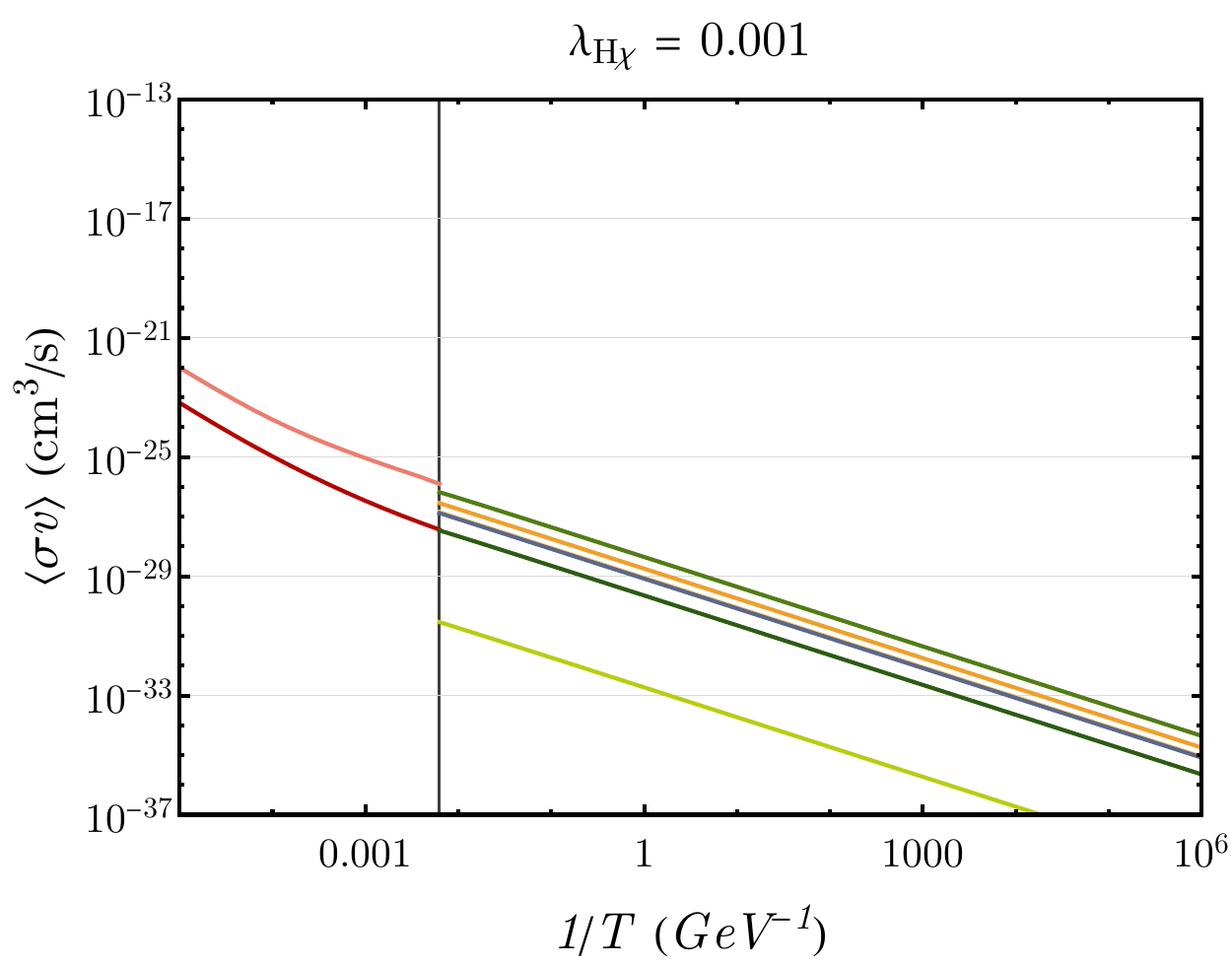}
	\includegraphics[width=0.32\linewidth]{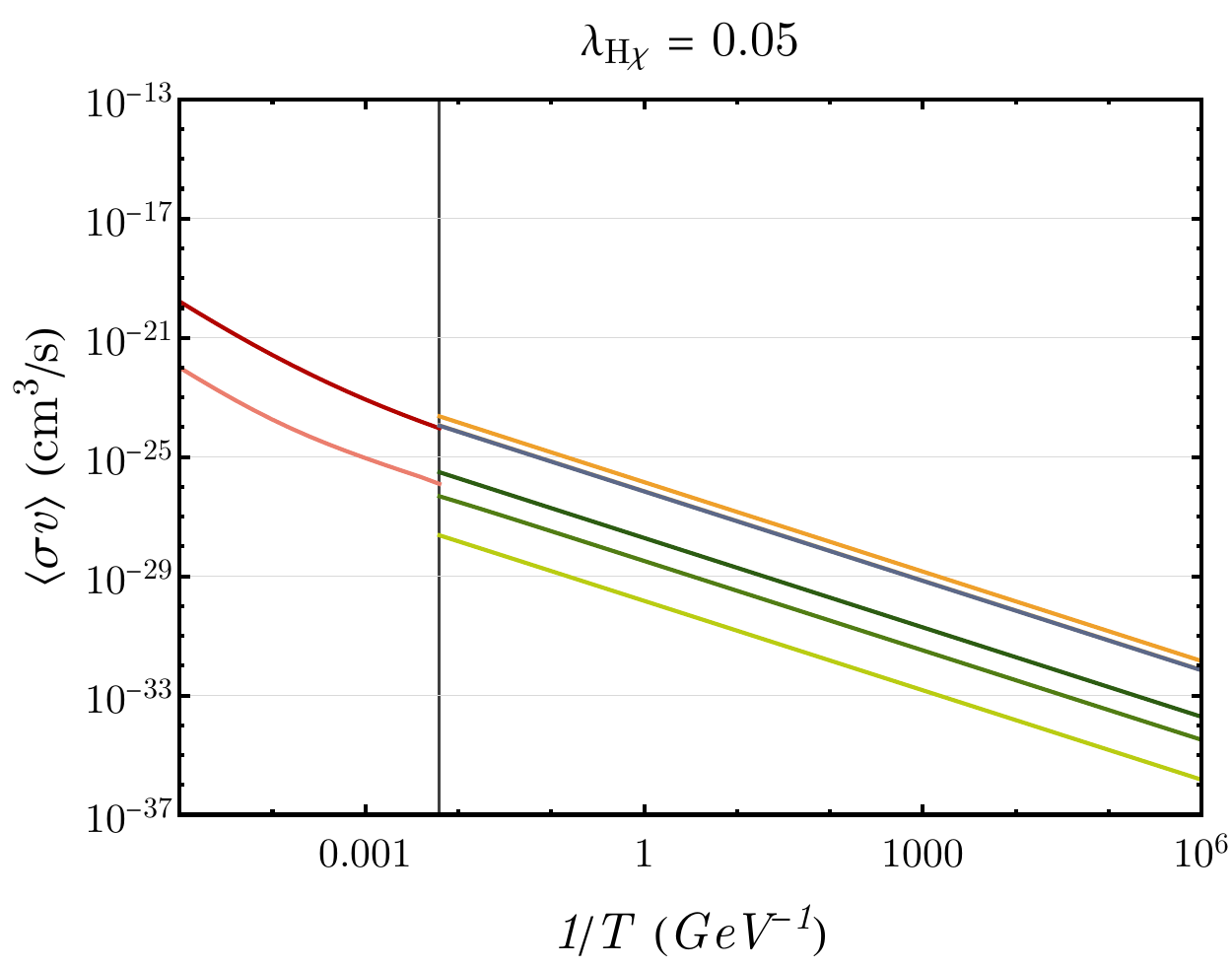}
	\includegraphics[width=0.32\linewidth]{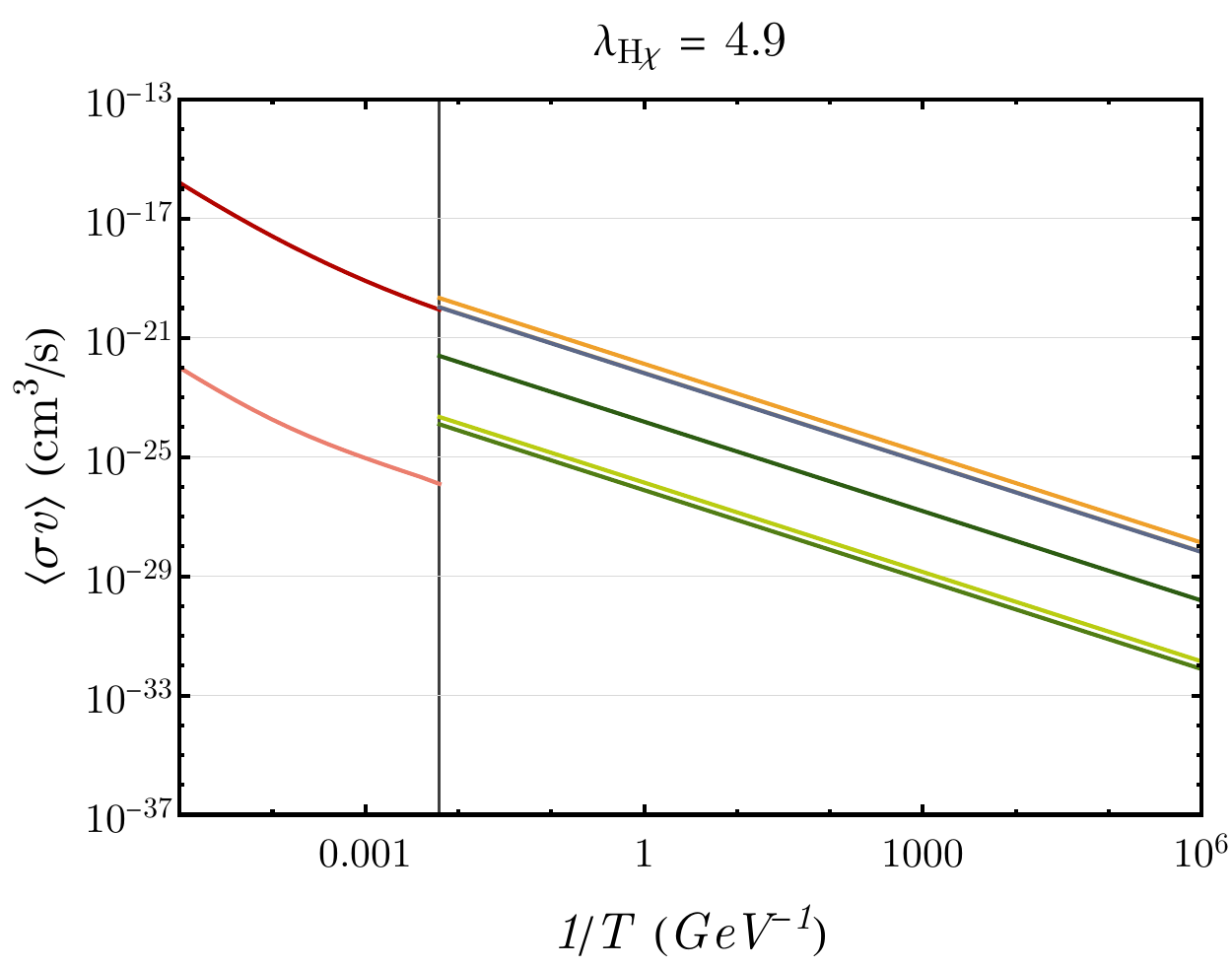}\\
	\includegraphics[width=0.9\linewidth]{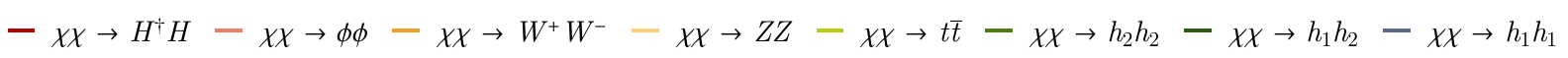}
	\caption{Contributions of individual annihilation channels to the \textit{improved} total thermal averaged cross-section, as a function of inverse temperature.}
	\label{fig:channels}
\end{figure}
For completeness, in Fig.~\ref{fig:channels} we also present the contributions of individual annihilation channels to the \textit{improved} total thermal averaged cross-section, for each benchmark point studied in Fig.~\ref{fig:vary_lamHC}.
As mentioned earlier, in the \textit{improved} approach and before EWSB, the annihilation channels contributing to $Y_\chi$ are $\chi \chi \to \phi \phi$, and $\chi \chi \to H H$.
When $\lambda_{\phi\chi} > \lambda_{H\chi}$, the total thermal averaged annihilation cross-section before EWSB is dominated by the process $\chi \chi \to \phi \phi$, which is in agreement with the first of this figure.
However, as $\lambda_{H\chi}$ is increased, we observe the channel $\chi \chi \to H H$ becoming the dominant one.

\section{Conclusions}
\label{sec:conclusion}

{\color{black} We propose a generic approach to calculate the DM relic density in TeV-scale Higgs-portal models, where the interplay between the occurrence of EWSB and the DM freeze-out instant is taken into account. In thermal freeze-out, one assumes that the DM particles remain in thermal equilibrium with the primordial bath until freeze-out occurs. In turn, this implies that the Boltzmann equation is solved by integrating from the freeze-out instant to the present time. In the standard approach, the DM particles and processes are always taken as those of the model's \emph{broken} phase. We have shown that in scenarios where DM freezes out \emph{before} EWSB, the use of the standard approach can lead to substantial discrepancies in the calculated value of the relic density.

It is important to stress that the effect we isolate here is 
\textbf{structural}: it results from the fact that the \emph{set of degrees of freedom and interaction channels} available to the DM is different in the unbroken and broken electroweak phases. This is conceptually distinct from (and independent of) finite-temperature radiative corrections to masses and couplings: the latter are perturbative refinements \emph{within} a given phase, whereas the former concerns \emph{which phase} applies at freeze-out. These two effects do not generically cancel, making TeV-scale Higgs-portal analyses in the literature that neglect both effects potentially incomplete in describing freeze-out near the electroweak transition. In this work, we focus exclusively on the phase-structure question and do not attempt a full finite-temperature resummed analysis, which constitutes a separate and model-dependent refinement.}

This has an important consequence for model building. By sticking to the standard approach, one can either exclude points in parameter space that are good, or allow points that in fact, do not yield the correct relic abundance. In the limit, one may exclude a good model or permit a wrong one. The \textit{improved} approach, described in our
study, makes this distinction:
we use the model in the unbroken phase, before EWSB;
and the model in the broken phase, after EWSB.
The differences between the two approaches have their origin in the thermally
averaged cross section, which is modified by EWSB with a redefinition
of masses and interactions.
We were able to quantify the difference between the calculations in the two scenarios
in a model independent way, in Eqs.~\eqref{eq:error}--\eqref{eq:beta1}.

We then made use of a simple model with two singlets and two independent symmetries; after EWSB, both the Higgs doublet and one of the scalars acquire vevs, breaking the $SU(2)\times U(1)$ and $Z_2^\prime$ symmetries, respectively.
We chose to work with this model and vacuum configuration
not only to simplify our analysis, but also because the differences between the 
\textit{standard} and \textit{improved} approaches were not that significant
in the (simpler) real scalar singlet extension of the SM.
Taking the DM mass above about 4~TeV after EWSB, we showed that indeed the two approaches lead to different relic densities.
Moreover, this difference can be either negative or positive,
implying the \textit{improved} relic density can lie below or above
the \textit{standard} relic density, respectively.

It is important to note that these heavy DM particles are not excluded by experiment. The ongoing and planned direct and indirect detection experiments (see  Ref.~\cite{Cirelli:2024ssz} for a comprehensive review) will most certainly
give rise to an increase (or even exclusion) in the upper bound of the DM mass in this simple scenario with two singlets and two independent symmetries. However, it is easy to extend the model with, for example, an extra singlet, such that the couplings responsible for direct and indirect detection are not the same as the ones that determine the value of the relic density. We defer a detailed investigation of such extensions to future work.

The study also builds a strong case for testing TeV-scale phenomena in upcoming high-energy particle colliders such as the FCC-hh~\cite{Benedikt:2022kan}.
In fact, although models for which our approach is important have very heavy DM candidates, they could in principle be probed at future colliders. In this case the approach could prove instrumental in the reconstruction of a given potential from an extended scalar sector.

\section{Acknowledgments}

This work is supported in part by the Portuguese
Fundação para a Ciência e Tecnologia (FCT) through the PRR (Recovery and Resilience
Plan), within the scope of the investment ``RE-C06-i06 - Science Plus Capacity Building",
measure ``RE-C06-i06.m02 - Reinforcement of financing for International Partnerships in Science,
Technology and Innovation of the PRR",
under the projects with references 2024.01362.CERN and 2024.03328.CERN.
The work of A.M. and J.P.S. is
also supported by FCT under Contracts UIDB/00777/2020, and UIDP/00777/2020.
The FCT projects are partially funded through
POCTI (FEDER), COMPETE, QREN, and the EU.
R.S. is partially supported by FCT under CFTC: UIDB/00618/2020
(https://doi.org/10.54499/UIDB/00618/2020), UIDP/00618/2020,
(https://doi.org/10.54499/UIDP/00618/2020).
A.M. is further supported by FCT with PhD Grant No. 2024.01340.BD.
SC thanks Genevieve Belanger for useful discussions and acknowledges support from the UKRI Future Leader Fellowship DarkMAP.


\appendix

\section{The potential after EWSB}
\label{app:Va}

The scalar potential in the unitary gauge after EWSB is given by
\begin{eqnarray}
V_{\rm aEWSB} &=& 
-\frac{1}{2}\mu_H^2 (v_H+h)^2 +	\frac{1}{4}\lambda_H \left( v_H+h  \right)^4 
-\frac{1}{2}\mu_\phi^2 (v_\phi+\phi)^2 +	\frac{1}{4!} \lambda_\phi \phi^4
\nonumber\\
&&+ \frac{1}{2}\mu_\chi^2 \chi^2 +	\frac{1}{4!} \lambda_\chi \chi^4
+ \frac{1}{4} \lambda_{H\phi} (v_H+h)^2 (v_\phi+\phi)^2 
\nonumber\\
&&+ \frac{1}{4} \lambda_{H\chi}(v_H+h)^2 \chi^2 
+ \frac{1}{4} \lambda_{\phi\chi} (v_\phi+\phi)^2 \chi^2.
\end{eqnarray}
The minimization of the potential after EWSB at the relevant extrema
implies the relations
\begin{eqnarray}
\frac{\partial V_{\rm aEWSB}}{\partial h} =0 
&\Leftrightarrow &
\mu_H^2 = v_H^2 \lambda_H + \frac{1}{2} v_\phi^2 \lambda_{H\phi},\\	
\frac{\partial V_{\rm aEWSB}}{\partial \phi} =0 
&\Leftrightarrow &
\mu_\phi^2 = \frac{1}{6} v_\phi^2 \lambda_\phi + \frac{1}{2} v_H^2 \lambda_{H\phi} .
\end{eqnarray}

To study the scalar mass spectrum after EWSB, we collect the quadratic terms in the fields:
\begin{equation}
V_{\rm aEWSB} \supset 
\frac{1}{2}
\begin{pmatrix}
h & \phi
\end{pmatrix}
\begin{pmatrix}
2 v_H^2 \lambda_H & v_H v_\phi \lambda_{H\phi}\\
v_H v_\phi \lambda_{H\phi} & \frac{1}{3} v_\phi^2 \lambda_\phi
\end{pmatrix}
\begin{pmatrix}
h \\
\phi
\end{pmatrix}
+
\frac{1}{2}
\left( \mu_\chi^2 + \frac{1}{2}v_H^2 \lambda_{H\chi} + \frac{1}{2}v_\phi^2 \lambda_{\phi\chi} \right) \chi^2.
\end{equation}
First, notice that this vacuum configuration allows $h$ and $\phi$ to mix.
To rotate these fields to the physical eigenbasis, we introduce $\theta$
through Eq.~\eqref{rot_theta}.
Then, the diagonalization of the squared mass matrix fixes the mixing angle at
\begin{equation}
	\tan\left( 2\theta \right) = 
	\frac{6\, v_H\, v_\phi\, \lambda_{H\phi}}
	{6 v_H^2 \lambda_H - v_\phi^2 \lambda_\phi},
\label{tan_theta}
\end{equation}
and yields the masses in Eqs.~\eqref{11}--\eqref{33}.
The external (input) parameters of this model are given in Eq.~\eqref{parameters_aEWSB},
while all the other parameters can be internally defined as
\begin{align}
v_\phi 
&= \frac{s_\theta c_\theta}{v_H \lambda_{H\phi}} \left( m_{h_1}^2 - m_{h_2}^2 \right),\\
\lambda_H 
&= \frac{1}{2 v_H^2}\left( m_{h_1}^2 c_\theta^2 + m_{h_2}^2 s_\theta^2 \right),\\
\lambda_\phi 
&= \frac{3}{v_\phi^2}\left( m_{h_1}^2 s_\theta^2 + m_{h_2}^2 c_\theta^2 \right),\\
\mu_H^2 
&= v_H^2 \lambda_H + \frac{1}{2} v_\phi^2 \lambda_{H\phi},\\
\mu_\phi^2 
&= \frac{1}{6} v_\phi^2 \lambda_\phi + \frac{1}{2} v_H^2 \lambda_{H\phi}, \\
\mu_\chi^2 
&=  m_\chi^2 - \frac{1}{2}v_H^2 \lambda_{H\chi} - \frac{1}{2}v_\phi^2 \lambda_{\phi\chi}.
\end{align}

The scalar mass spectrum before EWSB is related to the mass spectrum after EWSB through Eq.~\eqref{eq:masses_before}. Explicitely these read
\begin{eqnarray}
2\,\widetilde{m}_H^2 &=& 
\left( m_{h_1}^2 c_\theta^2 + m_{h_2}^2 s_\theta^2 \right) 
+ \frac{s_\theta^2 c_\theta^2}{v_H^2 \lambda_{H\phi}} 
\left( m_{h_1}^2 - m_{h_2}^2 \right)^2,\\
2\,\widetilde{m}_\phi^2 &=&
\left( m_{h_1}^2 s_\theta^2 
+ m_{h_2}^2 c_\theta^2 \right) + v_H^2 \lambda_{H\phi}, \\
\widetilde{m}_\chi^2 &=&
m_\chi^2 
- \frac{1}{2}v_H^2 \lambda_{H\chi} 
- \frac{1}{2}\frac{\lambda_{\phi\chi}}{\lambda_{H\phi}^2}
\frac{ s_\theta^2 c_\theta^2}{v_H^2 } \left( m_{h_1}^2 - m_{h_2}^2 \right)^2.
\end{eqnarray}
Notice that the values of the external parameters must be such that the scalar potential has a minimum both before and after EWSB. This condition is met when, before EWSB, the Hessian matrix is semi-definite, \textit{i.e.} when $\widetilde{m}_H^2$, $\widetilde{m}_\phi^2$ and $\widetilde{m}_\chi^2$ are positive.

\section{Boundedness from below}
\label{app:BFB}

For the scalar potential studied in this paper to be 
bounded from below, we must ensure that~\cite{Kannike:2012pe,Habibolahi:2022rcd}:
\begin{equation}
\lambda_H > 0\,, \quad\quad \lambda_{\phi} > 0\,, \quad\quad \lambda_{\chi} > 0\,.
\end{equation}
\begin{equation}
\lambda_{H \chi} + \sqrt{\frac{2}{3}\lambda_H \lambda_\chi} > 0\,, \quad \quad
\lambda_{H \phi} + \sqrt{\frac{2}{3}\lambda_H \lambda_\phi} > 0\,, \quad \quad
3 \lambda_{\phi \chi} + \sqrt{\lambda_\phi \lambda_\chi} > 0\,. 
\end{equation}
Then, when $3 \lambda_{\phi \chi} < \sqrt{\lambda_\chi  \lambda_\phi }$
and
$\lambda_{H\phi} < -  \lambda_{\phi\chi}\sqrt{6\lambda_H/\lambda_\chi}$
it is sufficient to require
\begin{equation}
9 \lambda_{H\phi}\lambda_{\phi\chi} - 3\lambda_{H\chi} \lambda_\phi 
    < \sqrt{
    3\left( 3 \lambda_{H\phi}^2 - 2\lambda_H \lambda_\phi \right)
    \left( 9 \lambda_{\phi\chi}^2 - \lambda_\phi \lambda_\chi \right)
    },
\end{equation}

\section{Direct Detection}
\label{app:direct}

The DM-nucleon spin-independent elastic 
scattering cross-section ($\sigma_{\mathrm{SI}}$),
in this model reads~\cite{Cerdeno:2010jj}
\begin{eqnarray}
\sigma_{\mathrm{SI}}
&=&
\frac{1}{\pi}
\left( \frac{m_N^2}{m_N + m_\chi} \right)^2
\frac{f_N^2}{v_H^2}
\left|
\frac{\cos \theta \lambda_{h_1 \chi \chi}}{m_{h_1}^2}
-\frac{\sin \theta \lambda_{h_2 \chi \chi}}{m_{h_2}^2}
\right|^2,
\end{eqnarray}
where
$f_N \approx 0.308$ is the Higgs-nucleon effective 
coupling~\cite{Hoferichter:2017olk},
$m_N \approx 0.938$\,GeV is the nucleon mass,
and
\begin{eqnarray}
\lambda_{h_1 \chi \chi} &=&
\cos \theta \, v_H \lambda_{H\chi} + \sin \theta \,  v_\phi \lambda_{\phi \chi},\\
\lambda_{h_2 \chi \chi} &=&
-\sin \theta \, v_H \lambda_{H\chi} + \cos \theta \, v_\phi \lambda_{\phi \chi}.
\end{eqnarray}

\end{document}